\documentclass[12pt]{article}
\usepackage{epsfig}
\usepackage{latexsym}
\usepackage{math}
\usepackage{graphicx}
\begin{document}
\parskip=4mm
\parindent=0cm
\newcommand{\Le}{{\mbox{\tiny\hspace{-0.4mm}$L$}}}
\newcommand{\Ri}{{\mbox{\tiny\hspace{-0.4mm}$R$}}}
\newcommand{\Ka}{{\mbox{\tiny\hspace{-0.4mm}$K$}}}
\title{Pion-Kaon Scattering near the Threshold in Chiral SU(2) Perturbation 
Theory}
\author{A. Roessl \\ {\small Institut de physique th\'{e}orique, Universit\'{e} de 
Lausanne, Switzerland}}
\date{}
\maketitle
\begin{abstract}
In the context of chiral $SU(2)$ perturbation theory, pion-kaon scattering is 
analysed near the threshold to fourth chiral order. The scattering amplitude 
is calculated both in the relativistic framework and by using an approach 
similar to heavy baryon chiral perturbation theory. Both methods lead to 
equivalent results. 
We obtain relations between 
threshold parameters, valid to fourth chiral order, 
where all those combinations of low-energy constants which are not associated 
with chiral-symmetry breaking terms drop out. The remaining low-energy constants 
can be estimated using chiral $SU(3)$ symmetry. Unfortunately, the experimental 
information is not precise enough to test our low-energy theorems.
\end{abstract}
\pagebreak

\pagebreak \section{Introduction}
By the end of the fifties it was clear that meson-meson scattering 
plays an important part in the understanding of the forces between baryons. 
For energies below the two-kaon threshold it is sufficient to take into 
account the interactions between pions. At higher energies, heavier 
degrees of freedom (kaon, eta, rho\dots) become relevant. Pion-kaon 
scattering -- with a threshold at about $650$ MeV -- constitutes the simplest 
mesonic system with unequal mass kinematics and carrying strangeness. For 
reviews of theoretical and experimental knowledge in the meson sector up to 
the late seventies see \cite{Petersen}, \cite{Martin}, \cite{Lang}. 

In the late seventies and early eighties, it was discovered that the low-energy 
structure of Green's functions of QCD can be determined systematically 
\cite{Weinberg, GaLe84, GaLe85}. The method -- called chiral 
perturbation theory ($\chi{\rm PT}$) -- is based on a simultaneous expansion of 
Green's functions in powers of external momenta and light quark masses, the 
latter taking into account the breaking of the $SU(N)_{\Ri} \times SU(N)_{\Le}$ (N 
flavour) 
symmetry of the strong interactions. In the context of $\chi SU(3)$ the 
expansion parameters $M_{\pi}^2 /(4\pi F_\pi)^2$ and $M_{\Ka}^2/(4\pi F_\pi)^2$ are 
of the same chiral order despite the fact that the pion mass $M_\pi$ is 
about three times smaller than the kaon mass $M_{\Ka}$. Chiral $SU(3)$ perturbation 
theory was applied to pion-kaon scattering in \cite{BKMthresh, MBK}. In these 
references the results for scattering lengths and other threshold parameters 
are given only numerically.

In 1987 $\chi{\rm PT}$ was extended to include nucleons, particles whose masses 
are non-zero in the chiral limit (of vanishing quark masses) \cite{GSS}. 
Interactions of these heavy degrees of freedom with pseudo-Goldstone bosons 
are not constrained by the full chiral group $G=SU(2)_{\Ri}  \times SU(2)_{\Le}$ but 
only by the diagonal subgroup $H=SU(2)_V$. There are therefore more unknown 
low-energy constants than there would be in the case of `Goldstones' only.

In recent years, several attempts have been made to extend the range of 
application of the $\chi SU(3)$ scattering amplitude to higher energies. In 
\cite{Hannah}, \cite{Beldjoudi} and \cite{Borges} this is achieved by using 
unitarisation methods and dispersion relations. Another possibility of 
partially including higher-order corrections is to take resonances 
into account explicitly \cite{BKMres}. 

An S-matrix parametrisation that conforms with unitarity \cite{Ishida} 
was applied to the data of two experiments on kaon-nucleon scattering \cite{Aston, 
Estabrooks} performed at SLAC in the seventies and eighties respectively. The authors 
find evidence for the existence of an $s$-wave resonance $\kappa(900)$, which 
constitutes an improvement in the understanding of $\pi$--$K$ partial waves. 
The light scalar mesons and their possible classifications are of much 
current interest (see for example \cite{Beveren}).

The purpose of the present article is to calculate the pion-kaon scattering 
amplitude in the context of 
$\chi SU(2)$ perturbation theory, where the strange quark mass is fixed by 
definition, to fouth chiral order and to find relations between threshold 
parameters where as much low-energy constants as possible cancel. An advantage 
of this approach is that the biggest expansion parameter (near the threshold) 
is now $M_{\pi}^2/M_{K}^2 <M_{\Ka}^2/(4\pi F_\pi)^2$ so that the chiral 
expansion is expected to converge more rapidly than in the $\chi SU(3)$ theory. 
However, analogously to baryon $\chi{\rm PT}$, the number of low-energy 
constants increases.

In section \ref{symm.sec} the chiral symmetry of the strong 
interactions is briefly reviewed. The role of closed kaon loops is discribed 
in section \ref{kta}. The transformation laws of the meson fields 
are given in section \ref{effF} and the general effective 
Lagrangian is constructed in section \ref{efflag}. 
Section \ref{masses.sec} is devoted to the chiral expansion of 
observables which can be extracted from the two point functions 
(pion and kaon masses, pion decay constant). In 
sections \ref{Methods.sec} and \ref{scatter}, 
we construct the isospin $3/2$ amplitude to fourth 
chiral order as a function of the low-energy constants. Section \ref{counting} 
concerns the matching of the $\chi SU(2)$ with the $\chi SU(3)$ amplitude and 
provides an expansion of the $\chi SU(2)$ coupling 
constants in powers of the strange quark mass. Low-energy theorems are given in 
section \ref{LET} and section \ref{summary} contains the conclusions.

\section{Chiral symmetry of the strong interactions}\label{symm.sec}
In the limit of vanishing light-quark masses ($m_u=m_d=0$) the Lagrangian of 
QCD, 
\begin{eqnarray}\label{QCD1}
{\cal L}^{\rm QCD} \hspace{-1mm}&=&\hspace{-1mm}-{1\over 2g^2} tr_c G_{\mu\nu}G^{\mu\nu} + {\rm i} \bar{q}_{\Ri}  
\gamma^\mu(\partial_\mu -{\rm i}G_\mu)q_{\Ri}  + {\rm i} \bar{q}_{\Le} 
\gamma^\mu(\partial_\mu -{\rm i}G_\mu)q_{\Le} \nonumber \\ 
&& \qquad -\bar{q}_{\Ri}  {\cal M} q_{\Le} - \bar{q}_{\Le} {\cal M} q_{\Ri}, 
\end{eqnarray}
is invariant under global chiral transformations \mbox{$(v_{\Ri}, v_{\Le}) 
\in G \hspace{-0.5mm}=\hspace{-0.5mm}SU(2)_{\Ri} 
\hspace{-0.5mm}\times\hspace{-0.5mm} SU(2)_{\Le}$}. The 
$3 \times 3$ colour matrix $G_\mu$ denotes the gluon field with associated field 
strength $G_{\mu \nu}$, and ${\cal M}={\rm diag}(m_u,m_d \dots)$ is the quark-mass 
matrix in flavour space. The action on the 
chiral up- and down-quarks,
\[q_{\Ri}  = {1\over 2}(1 + \gamma_5)q,\qquad q_{\Le} = {1\over 
2}(1 - \gamma_5)q, \] is defined by
\[q_{\Ri}  \rightarrow v_{\Ri}  q_{\Ri},\qquad q_{\Le} \rightarrow v_{\Le} 
q_{\Le}. \]
The gluon field and the heavy quarks $q_s$, $q_c$, $q_b$ and $q_t$ are not transformed 
under the chiral group. Therefore, these degrees of freedom are suppressed in 
the following. 
The vacuum of QCD is only assumed to be invariant under the isospin subgroup 
$H=SU(2)_V$ of $G$. According to Goldstone's 
theorem \cite{Goldstone} this implies the ocurrence of an internal $(N_G 
- N_H)$-dimensional subspace degenerate with the vacuum, the Goldstone bosons. 
Here $N_G$ and $N_H$ denote the number of parameters in the Lie groups $G$ and 
$H$ respectively (in our case $N_G=6$ and $N_H=3$).
The Green's functions of the conserved currents as well as the scalar and 
pseudoscalar densities are generated by the vacuum-to-vacuum amplitude
\[
{\rm e}^{{\rm i}Z(r,l,s,p)} = _{\rm out}\hspace{-1mm}\bra 0 \vert 0\ket_{\rm in}
\]
based on the Lagrangian
\begin{equation}\label{QCD2}
{\cal L}={\cal L}^{\rm QCD}_{{\cal M}=0} + \bar{q}_{\Ri}  r_\mu\gamma^\mu q_{\Ri}  + 
\bar{q}_{\Le} l_\mu\gamma^\mu q_{\Le} - \bar{q}_{\Ri} (s-{\rm i} p \gamma_5) 
q_{\Le} - \bar{q}_{\Le} (s-{\rm i} p \gamma_5) q_{\Ri}.                                          
\end{equation}
The external fields $r_\mu$, $l_\mu$, $s$ and $p$ are hermitian colour-neutral 
matrices in flavour space, $r_\mu$ and $l_\mu$ being traceless.
The Lagrangian (\ref{QCD2}) is now invariant under {\it local} chiral 
transformations provided the sources transform as
\begin{eqnarray}
r_\mu&\rightarrow& v_{\Ri}  r_\mu v_{\Ri} + {\rm i} v_{\Ri} \partial_\mu 
v_{\Ri}^+, \nonumber\\
l_\mu&\rightarrow& v_{\Le} l_\mu v_{\Le} + {\rm i} v_{\Le}\partial_\mu 
v_{\Le}^+, \nonumber\\
s+{\rm i}p&\rightarrow& v_{\Ri} (s + {\rm i}p) v_{\Le}. \nonumber
\end{eqnarray} 
Matrix elements of vector and axial vector currents 
\[V^\mu=\bar{q}\gamma^\mu\sigma^{i}q, \hspace{2cm} \qquad 
A^\mu=\bar{q}\gamma^\mu\gamma^5\sigma^{i}q \]
can be obtained by deriving the generating functional $Z$ with respect to 
\[\begin{array}{cc}
v_\mu^{i} = 
{1\over 2}(r_\mu^{i} + l_\mu^{i}) & \mbox{and  } a_\mu^{i} = 
{1\over 2}(r_\mu^{i} - l_\mu^{i})
\end{array}\]
where $r_\mu^{i}$ and $l_\mu^{i}$ are defined by $r_\mu=r_\mu^{i}\sigma^{i}$ 
and $l_\mu=l_\mu^{i}\sigma^{i}$.

\section{The role of closed kaon loops}\label{kta}

The role of closed nucleon loops is analysed in the context of baryon $\chi 
{\rm PT}$ in \cite{GSS}. The situation is 
analogue in the case of kaons instead of nucleons. The general effective Lagrangian describing the process $\pi K \rightarrow 
\pi K$ is given by a chirally symmetric string of terms
\begin{eqnarray}
{\cal L}\hspace{-1mm}&=&\hspace{-1mm}{\cal L}_{\pi \pi} + {\cal L}_{\pi K} 
\nonumber \\
{\cal L}_{\pi K} \hspace{-1mm}&=&\hspace{-1mm}{\cal L}_{K K} + 
{\cal L}_{K K K K} + \dots \nonumber
\end{eqnarray}
The term ${\cal L}_{\pi \pi}$ does not contain the kaon field $K$ whereas 
${\cal L}_{K K}$, ${\cal L}_{K K K K}$ \dots stand for local Lagrangians which 
are bilinear, quadrilinear, \dots in the kaon field. The bilinear term can be 
written in the form
\[
{\cal L}_{K K} = K^+ D K
\]
where $D$ denotes a $2 \times 2$ matrix-differential operator which contains 
only the pion fields $U$ and the external sources $a_\mu$, $v_\mu$, $s$ and $p$. The 
explicit form of $D$ is given in section \ref{Lagdens4.sub}. For brevity 
the fields $s$, $p$ and $v_\mu$ are dropped in the following. With an 
additional source $\rho$ for the kaon field,
\[
S(U,K,a,\rho)=\int {\rm d}^4 x \left( {\cal L}_{\pi \pi} + {\cal L}_{\pi K} + 
\rho^+ K + K^+ \rho \right)
\]
and the vacuum transition amplitude is given by
\begin{eqnarray}\label{pfadint}
_{\rm out}\hspace{-0.3 mm}\bra 0 \vert 0\ket_{\rm in} &=& {\rm e}^{{\rm i} 
Z(a,\rho,\rho^+)} \nonumber \\
&=& \int{\cal D}U {\cal D} K^+ {\cal D} K {\rm e}^{{\rm i} S(U, K, a, \rho)}.
\end{eqnarray}
In the effective theory with pions only n-loop 
contributions are suppressed by a factor of $O(q^{2n})$ \cite{Weinberg}. In the 
presence of heavy particles the situation is more complicated. First, 
there are closed kaon loops which are not suppressed. These loop contributions 
are real below the two-kaon threshold and, in a regularisation scheme which 
respects chiral symmetry, can be taken into account by a redefinition 
of fields and low-energy constants. This is always possible because -- 
by construction -- the general chiral Lagrangian 
${\cal L}$ contains all chirally invariant structures. In particular, 
as we are only interested in processes with one incoming and one outgoing kaon, 
vertices with more than two kaon legs can be dropped. In other words, we only consider Feynman 
diagrams with a single uninterrupted kaon line going through the diagram. 
The integration over the kaon fields can 
therefore be performed symbolically,
\begin{eqnarray}
{\rm e}^{{\rm i} Z(a,\rho,\rho^+)} 
&=& \int{\cal D}U {\cal D} K^+ {\cal D} K {\rm e}^{{\rm i} \int{{\rm d}^4 x 
\left( {\cal L}_{\pi \pi} \hspace{0.5mm}+\hspace{0.5mm} K^+ D K \hspace{0.5mm} 
+ \hspace{0.5mm}  
\rho^+ K + K^+ \rho \right) }} \nonumber \\
&=& {1 \over {\rm det}(D)} \int{\cal D}U {\rm e}^{{\rm i} \int{{\rm d}^4 x 
({\cal L}_{\pi \pi} \hspace{0.5mm}-\hspace{0.5mm} \rho^+ S_{\Ka}(U, a) \rho})},\nonumber 
\end{eqnarray}
if the low-energy constants are adjusted accordingly. 
Here $S_{\Ka}$ denotes the kaon propagator in the presence of 
pion and external fields, 
\[
D S_{\Ka}(x,y)= \delta^4(x-y).
\]
The above argument can be used again to justify the replacement 
of the determinant by $1$. It's only after this step that the low-energy 
constants contained in ${\cal L}_{\pi \pi}$ are rigorously the same as the 
ones in the original article by Gasser and Leutwyler \cite{GaLe84}.

Secondly, there are still unsuppressed loop contributions, arising from an 
internal kaon line, if the loop integrals are evaluated in the standard 
relativistic way. In the pion-nucleon case there are at least 
three different methods for evaluating such loop integrals in order to obtain 
the scattering amplitude to third chiral order. These methods, whose relationship has 
only been clarified recently \cite{Tang, Becher}, will 
be described in section \ref{Methods.sec}.

\section{Effective fields and transformation laws}\label{effF}
To model the dynamics of the pion-kaon system by an effective field theory, 
the action of the chiral group $G$ has to be chosen in such a way as to 
represent correctly the low-energy structure of QCD. The fact that the 
isospin subgroup leaves the vacuum invariant (in the chiral limit) means that 
$H$ must be the stability group (isotropy group) of each orbit of the 
carrier space $X$ containing the pions. This fact already implies that each 
orbit of $X$ must be homeomorphic to the coset space $G/H$ which is itself 
isomorphic to $S^3$ (the three-dimensional sphere) and also to 
$SU(2)$. It is convenient to select one representative 
element out of each equivalence class of $G/H$ defined by 
$(u_{\Ri} =u,u_{\Le}=u^+)$. This choice is unambiguous near the neutral element of 
$SU(2)$. The action of $(v_{\Ri} ,v_{\Le})\in G$ on $(u_{\Ri} ,u_{\Le})\in X$ is then:
\[
(u_{\Ri} ,u_{\Le}) \rightarrow (v_{\Ri} u_{\Ri} h^+,v_{\Le}u_{\Le}h^+)
\]
where the compensatory field $h\in SU(2)$ has to be chosen in such a way that 
$(v_{\Ri} u_{\Ri} h^+,v_{\Le}u_{\Le}h^+)$ is the representative element of the class containing 
$(v_{\Ri} u_{\Ri} ,v_{\Le}u_{\Le})$, i.e. so that $v_{\Ri} u h^+ = 
(v_{\Le}u^+ h^+)^+$. Because kaons have isospin ${1\over 2}$, this compensator 
can also be used to define the action of $G$ on the kaon fields 
$ K(x) \in \C^2 $:
\begin{equation}\label{chitrans}
\left[(u_{\Ri} ,u_{\Le});K\right] \rightarrow \left[(v_{\Ri} u_{\Ri} h^+,v_{\Le}u_{\Le}h^+);hK\right],
\end{equation}
so that in our case 
\begin{equation}\label{chitrans}
\left( u, K \right) \rightarrow \left( v_{\Ri}  u h^+, hK \right).
\end{equation}
The compensator $h$ does not depend on $u$ if the 
action is restricted to $H$ i.e. if $v_{\Ri} =v_{\Le}$. It was shown in a more general 
context by Coleman et al. \cite{CCWZ} that any action on the pair 
$(u,K)$ respecting the chiral symmetries given above is 
equivalent\footnote{Equivalence must be understood here to mean that the 
action (\ref{chitrans}) can be obtained by a field redefinition leaving the 
S-matrix unchanged.} to (\ref{chitrans}). In this work $u$ and $K$ are 
parameterised in terms of the physical particle fields as
\begin{equation}\label{mesonpar}
\begin{array}{ll}
u(x) = {\rm e}^{{\rm i}{\Phi \over 2F}}, & \Phi(x)=\left( \begin{array}{cc} 
\pi^0 (x) & -\sqrt{2} \pi^+ (x) \\ \sqrt{2} \pi^- (x) & -\pi^0 (x) 
\end{array}\right), \\
K(x)= \left( \begin{array}{c} K^+ (x) \\ K^0 (x) \end{array} \right), & 
K^+(x) = \left(  {\bar K}^0 (x), K^-(x)  \right).
\end{array}
\end{equation}

\subsection{Basic building blocks}\label{BBB}

The fields $U= u^2$ and $\chi=2 B (s+{\rm i}p)$ 
of standard chiral perturbation theory transform as $U \rightarrow v_{\Ri}  U v_{\Le}^{-1}$ 
under independent right and left chiral rotations $(v_{\Ri} ,v_{\Le}) \in SU(2) \times 
SU(2)$. 
In order to construct systematically all independent terms respecting the 
symmetries of the underlying theory, it is convenient to introduce fields $A$ 
which all transform in the same way under the chiral group. For the present 
application we choose
\begin{equation}\label{hah}
A \rightarrow hAh^+
\end{equation}
where $h$ is the compensator defined in section \ref{effF}. In this case the 
pions are contained in the matrix field $\Delta_{\mu}$ defined in table 
\ref{blocks}, which collects a set of basic building blocks sufficient for the 
construction of the fourth-order Lagrangian.
%
%
\begin{table}
\caption{{\small Definition and transformation of the basic building blocks under parity and charge 
conjugation. The space-time dependence of the fields is not shown. It is 
understood that the space arguments of parity-transformed fields undergo a 
change of sign. All the fields transform under chiral rotations as indicated in 
(\ref{hah}).}}
\begin{center} \begin{tabular}{|c|c|c|c|}
\hline
building block & definition & parity & charge conj.\\
\hline \hline
$\Delta_{\mu}$ & ${1\over2} u^+ D_\mu U u^+$ & $-\Delta^{\mu}$ & 
$(\Delta_{\mu})^T$ \\
\hline
$\Delta_{\mu\nu}$ & ${1\over 2}(D_\mu\Delta_\nu + D_\nu\Delta_\mu)$ & $-\Delta^{\mu\nu}$ & $(\Delta_{\mu\nu})^T$ \\ 
\hline
$KK^+$& & $KK^+$ & $(KK^+)^T$\\
\hline
$K_{\pm}^{\mu}$ & $D^{\mu}KK^+ \pm KD^{\mu}K^+$ & $K_{\mu\pm}$ & $\pm (K_{\pm}^{\mu})^T$ \\ 
\hline
$K_{\pm}^{\mu\nu}$ & $D^{\mu}KD^{\nu}K^+ \pm D^{\nu}KD^{\mu}K^+$ & $K_{\mu\nu\pm }$ & $\pm (K_{\pm}^{\mu\nu})^T$\\
\hline
$K^{*\mu\nu}_{\pm}$ & $D^{\mu\nu}KK^+ \pm KD^{\mu\nu}K^+$ & $K^{*}_{\mu\nu\pm}$ & $\pm (K^{*\mu\nu}_{\pm})^T$\\ 
\hline
$K_{\pm}^{\mu\nu\rho}$ & $D^{\mu\nu}KD^{\rho}K^+ \pm D^{\rho}KD^{\mu\nu}K^+$ & $K_{\mu\nu\rho\pm}$ & $\pm (K_{\pm}^{\mu\nu\rho})^T$\\
\hline 
$K^{\mu\nu\rho\sigma}_{\pm}$ & $D^{\mu\nu}KD^{\rho\sigma}K^+ \pm D^{\rho\sigma}KD^{\mu\nu}K^+$ & $K_{\mu\nu\rho\sigma\pm}$ & $\pm (K^{\mu\nu\rho\sigma}_{\pm})^T$\\ 
\hline
$\chi_{\pm}$ & $u^+ \chi u^+ \pm u \chi^+u$ &$\pm \chi_{\pm}$ & $\chi_{\pm}^T$\\
\hline 
$\chi^{\mu}_{\pm}$ & $u^+ D^{\mu}\chi u^+ \pm u D^{\mu}\chi^+u$ & $\pm \chi_{\mu\pm}$ & $(\chi^{\mu}_{\pm})^T$\\ 
\hline
$F^{\mu\nu}_{\pm}$ & $u^+ r^{\mu\nu}u \pm u l^{\mu\nu}u^+$ & $\pm F_{\mu\nu\pm}$ & $\mp (F^{\mu\nu}_{\pm})^T$ \\
\hline
$F^{\mu\nu\rho}_{\pm}$ & $u^+ D^{\mu}r^{\nu\rho}u \pm u D^{\mu}l^{\nu\rho}u^+$ & $\pm F_{\mu\nu\rho\pm}$ & $ \mp (F^{\mu\nu\rho}_{\pm})^T$\\
\hline
\end{tabular} \end{center}
\label{blocks}
\end{table}
The symbol $D^{\mu\nu}$ is defined by $D^{\mu\nu} = D^{\mu}D^{\nu} + 
D^{\nu}D^{\mu}$ where $D_{\mu}$ is the covariant derivative, transforming like the 
field on which it acts. For example
\begin{eqnarray}\label{Gamma}
D_{\mu} U &=& \partial_{\mu} U - {\rm i} r_{\mu} U + {\rm i} U l_{\mu} \nonumber \\
D_{\mu} K &=& \partial_{\mu} K + \Gamma_{\mu} K, \nonumber \\
\Gamma_{\mu} &=& {1 \over 2} \left( u^+ (\partial_{\mu} -{\rm i} r_{\mu}) u + 
u (\partial_{\mu} -{\rm i} l_{\mu} ) u^+ \right).
\end{eqnarray}

Kaon matrices like $KK^+$, which are defined by
\[
(KK^+)_{ij} = K_i K_j^+,
\]
are introduced in order to have the same transformation laws for all fields 
under the chiral group and so that trace relations can be used to eliminate 
redundant terms. In fact, for any set $A_1,A_2,A_3,A_4$ of complex $2\times 2$ 
matrices, 
{\footnotesize \begin{eqnarray}
\sum_{2p} \bra A_1A_2A_3 \ket \hspace{-2mm}&=&\hspace{-2mm} \sum_{3p} \bra A_1A_2 \ket \bra A_3 \ket - 
\bra A_1\ket \bra A_2\ket \bra A_3\ket, \nonumber \\ 
\sum_{6p} \bra A_1A_2A_3A_4 \ket \hspace{-2mm}&=&\hspace{-2mm} \sum_{3p} \bra A_1A_2 \ket \bra A_3A_4 
\ket + \sum_{6p} \bra A_1A_2\ket \bra A_3\ket \bra A_4\ket - 3 \bra A_1 \ket \bra A_2\ket 
\bra A_3\ket \bra A_4\ket. \nonumber
\end{eqnarray}}
\unskip The sum $\d\sum_{np}$ is taken over all $n$ permutations giving different 
expressions. For example
\[ \sum_{3p} \bra A_1A_2 \ket \bra A_3 \ket = \bra A_1A_2 \ket \bra A_3 \ket + 
\bra A_2A_3 \ket \bra A_1 \ket + \bra A_3A_1 \ket \bra A_2 \ket .\]

The differences $D_\mu D_\nu - D_\nu D_\mu$ and 
$D_\mu\Delta_\nu - D_\nu\Delta_\mu$ do not appear in the 
list of basic building blocks: due to the relations
\begin{eqnarray}
(D^{\mu}D^{\nu}-D^{\nu}D^{\mu})A &=& (-[\Delta^{\mu},\Delta^{\nu}] -{{\rm i}
\over{2}}F^{\mu\nu}_+ )A, \nonumber \\
D^\mu\Delta^\nu - D^\nu\Delta^\mu &=& -{{\rm i}\over 2} F_-^{\mu\nu}, \nonumber
\end{eqnarray}
such terms are contained in the set involving the fields 
$\Delta^{\mu}$ and $F^{\mu\nu}_{\pm}$.

Terms with three or more covariant derivatives on the field $K$ can be 
transformed by the total derivative argument described for example in 
\cite{Scherer} to give terms with two covariant derivatives only.
Table \ref{blocks} also shows the transformation properties under parity and 
charge conjugation.
%

\section{Construction of the effective Lagrangian to fourth chiral 
order}\label{efflag}
Our aim is to construct an effective field theory describing pion-kaon 
scattering in the energy region where spatial momenta of the mesons in the 
$CM$ frame are small compared to the heavy mass $M_{\Ka}$:
\[
q < M_\pi \sim 140  \hspace{1mm}{\rm MeV} < M_{\Ka} \sim 490  \hspace{1mm}{\rm MeV}.
\]
The theory should be invariant under the Lorentz group, parity and charge 
conjugation and should represent correctly the chiral-symmetry breaking 
structure. We consider the spatial 
momenta and the pion mass to be of $O(p^1$) whereas the kaon mass is 
considered to be of $O(p^0)$. 
Chiral-power counting is non-trivial in chiral perturbation theory if there 
are particles whose masses do not vanish in the chiral limit \cite{GSS}. Loops 
of these {\it heavy} particles have unsuppressed contributions if they are 
evaluated using the standard relativistic regularisation \cite{GaLe85}. 
If the heavy particle is a kaon, the kinetic term giving rise to the propagator 
$(M^2-p^2)^{-1}$ is 
\begin{equation}\label{ackin}
S_{\rm kin}=\int {\rm d}^4 x (\partial_\mu K^+ \partial^\mu K - M^2 K^+ K ). 
\end{equation}
In ${\rm HB}\chi{\rm PT}$ a consistent chiral-power counting procedure is obtained by a field 
redefinition \cite{JM}, which in the kaon case takes the form
\begin{equation}\label{nonreltrans}
K(x)={\rm e}^{{\rm i}Mvx} k(x),
\end{equation}
where $v$ is an arbitrary light-like four-component vector with $v^2=1$. 
In this way the mass term in equation (\ref{ackin}) is eliminated,
\[
S_{\rm kin}=\int {\rm d}^4 x (-2{\rm i}M v^\mu k^+ \partial_\mu k + \partial_\mu 
k^+ \partial^\mu k),
\]
and the propagator becomes $(-2Mvp)^{-1}$. The heavy-particle mass appears 
only as a global factor and can be taken in front of the loop integral, 
whereas the term $\partial_\mu k^+ \partial^\mu k$ must be treated as a 
vertex of second order. 

Equivalently, the heavy-particle expansion can be discussed on the level of 
Feynman diagrams. In this language, the relativistic propagator $(M^2-p^2)^{-
1}$ is developed in powers of the light momentum $l_\mu$ defined by 
$p_\mu=Mv_\mu+l_\mu$ :
\[
{1\over M^2-(Mv+l)^2}
= -{1\over 2Mvl} +{l^2\over 4M^2(vl)^2} + O(l^1).
\]
The heavy mass again factorises out and integrals with a single uninterupted 
kaon line are powers -- given by dimensional considerations -- of the light 
scales only. It is shown explicitly in appendix \ref{rel.ap} that the 
development of the kaon propagator under the integral sign does not lead to 
inconsistencies in the application to elastic $\pi$--$K$ scattering. 

The above discussion illustrates that chiral-power counting at the Lagrange level is 
easier in the `non-relativistic' formalism: terms with one $\partial_\mu$ just 
count as $O(p^1)$ and $\chi_{\pm}$ as $O(p^2)$. To 
obtain the Lagrange density to fourth chiral order, we therefore write down in 
a first step all invariant non-relativistic structures. These terms are given 
in appendix \ref{nonrelstr}. In a second step, we determine a set of 
relativistic terms, which generate -- via the relation (\ref{nonreltrans}) -- 
those non-relativistic ones.

\subsection{Lagrange density to fourth order}\label{Lagdens4.sub}
The effective $\chi SU(2)$ Lagrangian contributing to the pion-kaon scattering 
amplitude up to fourth chiral order is given by
\begin{eqnarray}\label{Lagrangian}
{\cal L} &=& {\cal L}^{(2)}_{\pi\pi} + {\cal L}_{\pi\pi}^{(4)} + 
{\cal L}_{\pi K}^{(1)} + {\cal L}_{\pi K}^{(2)} + {\cal L}_{\pi K}^{(3)} +
{\cal L}_{\pi K}^{(4)}, \nonumber \\
\nonumber \\
{\cal L}_{\pi\pi}^{(2)} &=& F^2 \left( -\bra\Delta_\mu\Delta^\mu\ket +
{1\over 4} \bra \chi_+ \ket \right), \nonumber \\
\nonumber \\
{\cal L}_{\pi\pi}^{(4)} &=& 4 l_1 \bra\Delta_\mu\Delta^\mu\ket^2 + 
4 l_2 \bra\Delta_\mu\Delta_\nu\ket\bra\Delta^\mu\Delta^\nu\ket \nonumber \\
&&+ {1\over 16} (l_3 + l_4) \bra \chi_+ \ket^2 
- {1\over 2} l_4 \bra\Delta_\mu\Delta^\mu\ket\bra\bar{\chi}_+\ket, \nonumber \\
\nonumber \\
{\cal L}_{\pi K}^{(1)} &=& D_\mu K^+D^\mu K -M^2 K^+K, \nonumber \\
\nonumber \\
{\cal L}_{\pi K}^{(2)} &=&
A_1\;\bra\Delta_\mu\Delta^\mu\ket K^+K \nonumber \\
&&+ A_2\;\bra\Delta^\mu\Delta^\nu\ket D_\mu K^+D_\nu K \nonumber \\
&&+ A_3\;K^+\chi_+K \nonumber \\
&&+ A_4\;\bra \chi_+\ket K^+K, \nonumber \\
\nonumber \\
{\cal L}_{\pi K}^{(3)} &=& B_1 \left( K^+\left[\Delta^{\nu\mu},\Delta_\nu\right]D_\mu K-D_\mu 
            K^+\left[\Delta^{\nu\mu},\Delta_\nu\right]K \right) \nonumber \\ 
&&+ B_2\;\bra\Delta^{\mu\nu} \Delta^\rho\ket 
\left(D_{\mu\nu} K^+D_\rho K + D_\rho K^+ D_{\mu\nu}K \right) \nonumber \\
&&+ B_3\; \left( K^+\left[\Delta_\mu,\chi_-\right]D^\mu K-D_\mu 
K^+\left[\Delta^\mu,\chi_-\right]K \right),
\end{eqnarray}
\begin{eqnarray}
{\cal L}_{\pi K}^{(4)} &=&
C_1\;\bra\Delta_\nu \Delta^{\mu\nu}\ket \left(K^+D_\mu K+D_\mu 
            K^+K\right) \nonumber \\
&&+ C_2\;\bra\Delta^{\mu\rho} \Delta^\nu\ket 
\left(D_{\mu\nu} K^+D_\rho K + D_\rho K^+ D_{\mu\nu}K\right) \nonumber \\
&&+ C_3\;\left( \bra\Delta^{\mu\nu} \Delta^\rho\ket 
\left(D_{\mu\nu} K^+D_\rho K + D_\rho K^+ D_{\mu\nu}K \right) \right. 
\nonumber \\
&& \left. -2 \left( D^{\mu\nu} K^+ \Delta_\mu\Delta_{\nu\rho} D^\rho K + 
D^\rho K^+\Delta_{\nu\rho}\Delta_\mu D^{\mu\nu} K \right) \right) \nonumber \\
&&+ C_4\;\bra\Delta^{\mu\nu} \Delta^{\rho\sigma}\ket 
\left(D_{\mu\nu} K^+D_{\rho\sigma} K + D_{\rho\sigma} K^+ D_{\mu\nu}K\right)
\nonumber \\
&&+ C_5\; \left( D_{\mu}K^+\chi_+D^{\mu}K - M^2 K^+\chi_+K \right) \nonumber \\
&&+ C_6\; \left( \bra \chi_+\ket D_\mu K^+ D^\mu K-M^2 \bra \chi_+\ket K^+K 
\right) \nonumber \\
&&+ C_7\; \bra\Delta_\mu\chi_-\ket\left(K^+ D_\mu K+D_\mu 
K^+K\right) \nonumber \\
&&+ C_8\; \bra\Delta_\mu\Delta^\mu\ket K^+ \chi_+K \nonumber \\
&&+ C_9\; \bra\Delta_\mu\Delta^\mu\ket\bra\chi_+\ket K^+K 
\nonumber \\  
&&+ C_{10}\; \bra\Delta^\mu\Delta^\nu\ket \left( D_\mu K^+ \chi_+
D_\nu K + D_\nu K^+ \chi_+ D_\mu K \right) \nonumber \\  
&&+ C_{11}\; \bra\Delta^\mu\Delta^\nu\ket \bra \chi_+ \ket
\left( D_\mu K^+ D_\nu K + D_\nu K^+ D_\mu K \right) \nonumber \\  
&&+ C_{12}\; D_\mu K^+ \{ \{\Delta^\mu,\Delta^\nu \}, \chi_+ \}  D_\nu K
\nonumber \\  
&&+ C_{13}\; \bra \chi_+\ket K^+\chi_+K \nonumber \\
&&+ C_{14}\; \bra \chi_+^2\ket K^+K \nonumber \\
&&+ C_{15}\; \bra \chi_+\ket^2 K^+K \nonumber \\
&&+ C_{16}\; \bra \chi_-^2\ket K^+K. \nonumber
\end{eqnarray}

The chiral order associated with each class of terms corresponds to the 
dominating non-relativistic structures generated from the relativistic terms, 
or equivalently, to the chiral order of the leading tree-contributions. 

The constants $l_i$ in the pionic part of the Lagrangian of fourth order are 
chosen in such a way that they coincide with the original ones first 
introduced in \cite{GaLe84}, where a different parameterisation is used for the 
unitary matrix $U$. Here, by convention, all the terms involving 
$K^{*,\mu\nu}_{\pm}, K^{\mu\nu\rho}_{\pm}, K^{\mu\nu\rho\sigma}_{\pm}, 
\chi_\mu$ and $F_{\pm}^{\mu\nu\rho}$ are eliminated by shifting the 
covariant derivatives to other fields, thus creating equation-of-motion terms, 
total derivatives or other basic building blocks. The elimination of the field 
$\chi_\mu$, which is used in \cite{GaLe84}, explains the appearence of the 
combination $l_3 + l_4$ in the third term of ${\cal L}^{(4)}_{\pi \pi}$.

Some chiral-symmetry breaking terms are proportional to each other in the isospin 
limit ($m_u=m_d$). They are, however, distinguished for reasons of generality.
 
\subsection{The classical equations of motion}
The classical equations of motion associated with the fourth order Lagrangian 
are given by
\begin{eqnarray}
4F^2 D_\mu\Delta^\mu &=& F^2\chi_- 
-{1\over2} F^2 \bra \chi_- \ket + O(K), \nonumber \\
D_\mu D^\mu K + M^2 K &=& O(p^2).\nonumber
\end{eqnarray}

The explicit form of the higher-order terms is not needed in the present approach. 
Here the only use of these equations is to eliminate interaction terms 
proportional to $D_{\mu} \Delta^{\mu}$ or $D_\mu D^\mu K$ by appropriate field 
redefinitions \cite{Scherer}.
\section{Masses and the pion decay constant}\label{masses.sec}
Using the definitions for the effective fields $\Delta_\mu$, $\Gamma_\mu$, 
$\chi_+$ and $\chi_-$ given in section \ref{effF}, one can turn off all 
external sources and develop the general Lagrange density in powers of the 
pion and kaon fields:
{\small \begin{eqnarray}
{\cal L}&=&{1 \over 4} (1+2 {m_0^2\over F^2} l_4) \bra \partial_\mu \Phi 
\partial^\mu \Phi \ket + {m_0^2 \over 4} \left[1+ 2 {m_0^2\over F^2} (l_3+l_4) 
\right] \bra \Phi 
\Phi \ket \nonumber \\
&&\hspace{2mm}+ \left[ 1+ 2 m_0^2(C_5 +2 C_6)\right] \partial_\mu K^+ \partial^\mu K 
\nonumber \\
&&\hspace{4mm}+ \left[ M^2-2 m_0^2(A_3+2 A_4-M^2 (C_5 + 2 C_6))+O(m_0^4)\right] K^+ K \nonumber \\
&&\hspace{6mm}+ O(\Phi^4,\Phi^2 K^2),\nonumber
\end{eqnarray}}
where $m_0^2 = 2B{\hat m}$ and ${\hat m}=m_u=m_d$.

\begin{figure}
\begin{center}
\epsfig{file=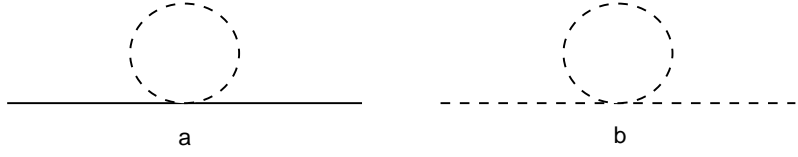,width=.75\textwidth}
\end{center}
\vspace{-1cm}
\caption{{\small Loop contributions to the kaon (a) and pion (b) two-point functions. 
The continuous and dashed lines stand for kaon and pion propagators 
respectively. The relevant interactions are ${\cal L}^{(2)}_{\pi K}$ for the 
kaon and ${\cal L}^{(2)}_{\pi \pi}$ for the pion. Vertices from 
${\cal L}^{(1)}_{\pi K}$ do not contribute. As explained in section \ref{kta}, 
closed kaon-loops must not be taken into account.}}\label{twopoint.fig}
\end{figure}

The field redefinitions
\begin{eqnarray}
\Phi '&=& \left( 1 + {m_0^2\over F^2} l_4 \right) \Phi, \nonumber\\
K '&=& \left( 1+ m_0^2(C_5 +2 C_6) \right) K, \nonumber
\end{eqnarray}
give the standard form for the kinetic terms in the Lagrangian,
\begin{eqnarray}
{\cal L}&=&{1\over 4} \left( \bra \partial_\mu \Phi ' \partial^\mu \Phi ' \ket + 
M_{\pi,{\rm t}}^2 \bra \Phi ' \Phi ' \ket \right) \nonumber \\
&&\hspace{2mm}+ \left( \partial_\mu K'^+ \partial^\mu K' + M_{K,{\rm t}}^2 K'^+ K' \right) 
+ O(\Phi^4,\Phi^2 K^2),\nonumber
\end{eqnarray}
where $M_{\pi,{\rm t}}^2=m_0^2(1+2 l_3 m_0^2/F^2)$ and 
$M_{K,{\rm t}}^2=M^2-2 m_0^2 (A_3 + 2 A_4) + O(m_0^4)$ are the masses at tree level. Including 
the pion one-loop contributions shown in figure \ref{twopoint.fig}, the 
physical pion and kaon masses to fourth chiral order are 
\begin{eqnarray}\label{mpimk}
M_\pi^2 &=& m_0^2 \left( 1 +2{m_0^2\over F^2}l_3^r + {m_0^2\over F^2}{\bar 
\mu}_\pi \right), \nonumber \\ 
M_{\Ka}^2 &=& M^2 + M^{(2)} m_0^2 + M^{(4)} m_0^4, 
\end{eqnarray}
where
\begin{eqnarray}
M^{(2)}&=&-2 \sigma_0, \nonumber \\
M^{(4)}&=& 4 \sigma_0 (C_5^r +2 C_6^r)-8(C_{13}^r+C_{14}^r+2C_{15}^r)-{3 M^2 
A_2 \over 256 \pi^2 F_{\pi}^2} \nonumber \\ 
&&+{\bar \mu}_{\pi}\left( {3 
A_1 \over F_{\pi}^2 }+{6 \sigma_0 \over F_{\pi}^2 }+{3 M^2 A_2 \over 
4 F_{\pi}^2 } \right), \nonumber \\
\sigma_0&=&A_3+2 A_4, \nonumber \\
{\bar \mu}_\pi &=& { 1 \over 32 \pi^2} \ln {m_0^2 \over \mu^2}.\nonumber
\end{eqnarray}
Since the first two coefficients in the kaon-mass expansion are finite, the 
constants $M$ and $\sigma_0$ are not renormalised.

The pion decay constant has been calculated in \cite{GaLe84} to fourth chiral 
order as
\begin{equation}\label{pidec.const}
F_{\pi} = F \left(1 + {m_0^2\over F^2} l_4^r -2 {m_0^2 \over F^2} {\bar 
\mu}_\pi \right).
\end{equation}
This result is not modified because the effect of closed kaon loops is 
already included in the $\chi SU(2)$ coupling constants.

The {\it kaon} decay constant cannot be obtained in the present framework, 
because strangeness-changing, axial currents have not been coupled to 
external sources in the general $\chi SU(2)$ Lagrangian. 

\subsection{Matching with chiral $SU(3)$}

In the $\chi SU(3)$ theory the strange-quark mass is treated as an expansion 
pa\-ra\-me\-ter, whereas in the $\chi SU(2)$ case it is fixed and contained 
in the low-energy constants, just like all the other heavy-quark 
masses. Comparing the expressions for $M_{\pi}$, $M_{\Ka}$ and $F_{\pi}$, given by 
(\ref{mpimk}) and (\ref{pidec.const}) respectively, with 
the corresponding $\chi SU(3)$ results \cite{GaLe85} leads to expansions of the 
low-energy constants in powers of the strange-quark mass (${\bar M}_K^2=B_0 
m_s$):
{\small \begin{eqnarray}
M^2 \hspace{-2mm}&=&\hspace{-2mm} {\bar M}_K^2 +{ {\bar M}_K^4 \over F_0^2} \left[ { 1 \over 36 \pi^2} 
\ln({4 {\bar M}_K^2 \over 3 \mu^2}) - 8(2L_4^r +L_5^r - 4L_6^r-2L_8^r) \right] 
\hspace{-1mm}+ O({\bar M}_K^6), \nonumber \\
\nonumber\\
\sigma_0 \hspace{-2mm}&=&\hspace{-2mm} -{1\over 4} - { {\bar M}_K^2 
\over F_0^2} \left[ {1\over 288 \pi^2} + { 1 \over 72 \pi^2}\ln({4 {\bar M}_K^2 
\over 3 \mu^2}) - 4(2L_4^r +L_5^r - 4L_6^r-2L_8^r) \right] \hspace{-1mm} + O({\bar M}_K^4), 
\nonumber \\
&& \hspace{-1.3cm}{F^2\over 
8}(C_5^r+2C_6^r)+F^2(C_{13}^r+C_{14}^r+C_{15}^r) + {3 M^2 A_2 \over 2048 \pi^2} \nonumber
 \\
&=&\hspace{-1mm} -{5 \over 9216 \pi^2}-{1 \over 2304 \pi^2} \ln({4 {\bar M}_K^2 \over 3 \mu^2})
+{1\over 4}(4L_4^r +L_5^r - 8L_6^r-2L_8^r) + O({\bar M}_K^2). \nonumber
\end{eqnarray}}

The combination of low-energy constants associated with the second order Lagrangian 
${\cal L}_{\pi K}^{(2)}$, 
\[
A_1 +{M^2 A_2 \over 4} = {1\over 2}+O({\bar M}_K^2),
\]
is evaluated to next-to-leading order in section (\ref{counting}).

The results for the expansions of pionic low-energy constants coincide with 
those given in \cite{GaLe85} and are not repeated here.

\section{Relativistic versus non-relativistic loops}\label{Methods.sec}
  Chiral-power counting is a non-trivial problem whenever particles are present 
whose masses do not vanish in the chiral limit. In baryon $\chi{\rm PT}$ there 
are three main approaches to the calculation of the scattering amplitude to 
third chiral order \cite{GSS, Tang, MartinM}. The relativistic 
approach \cite{GSS} maintains relativistic invariance during the whole calculation 
and uses the same techniques of dimensional regularisation that are used in 
standard $\chi{\rm PT}$ with (pseudo-) Goldstone bosons only. The chiral power $D$ of 
the non-polynomial part of a given loop diagram without closed nucleon loops 
is given by 
\begin{equation}\label{chidim}
D=2 L + 1+\sum_d (d-2) N^M_d +\sum_d (d-1) N^{M N}_d
\end{equation}
where $L$ is the number of loops and $N^M_d$($N^{M N}_d$) the number of pure 
meson (meson-nucleon) vertices, whose tree contributions are of chiral order 
$d$. However, loop diagrams usually contain also terms of lower order than 
$D$. 
Although these terms can be absorbed into the low-energy 
constants, most applications to pion-nucleon scattering are performed 
in the `non-relativistic' approach (${\rm HB}\chi{\rm PT}$).  

In ${\rm HB}\chi{\rm PT}$ \cite{MartinM} the Lorentz-invariant Lagrange 
density is written as a sum of terms, which break in general Lorentz 
invariance, and loop calculations are performed using these non-invariant 
vertices. In this approach the spin structure is simplified and the power-counting 
formula (\ref{chidim}) can easily be proved. 
Lorentz invariance can be recovered at a later stage but there 
are many more vertices than in the relativistic case. Furthermore 
${\rm HB}\chi{\rm PT}$ suffers from a deficiency: The perturbation series 
corresponding to the non-relativistic expansion fails to converge in part of 
the low-energy region \cite{Becher}.

An alternative to the above method is used in \cite{Tang}, 
where the heavy-particle expansion is used only under the loop integral and 
not -- as in ${\rm HB}\chi{\rm PT}$ -- on the Lagrangian level. This method is 
relativistic at every stage of the calculation and the power-counting formula 
(\ref{chidim}) is also valid for the divergent part. Although this method 
suffers in general from the same deficiency as ${\rm HB}\chi{\rm PT}$ it 
seems to be the most efficient one for pion-kaon scattering to one loop 
because in that case -- due to the absence of cubic 
interaction terms -- there are no such low-energy divergencies as those 
mentioned above in the case of ${\rm HB}\chi{\rm PT}$. This is shown 
explicitely in appendix \ref{rel.ap}.

\section{The scattering amplitude}\label{scatter}

The isospin $3/2$ amplitude is defined by
\[ 
_{\rm out}\bra\pi^+(\vec{p}_3)K^+(\vec{p}_4)|\pi^+(\vec{p}_1)K^+(\vec{p}_2)\ket_{\rm 
in}={\rm i}(2 \pi)^4\delta^{(4)}(p_1+p_2-p_3-p_4)T^{3\over 
2}(s,t,u). \]

where $s,u$ and $t$ are the conventional Mandelstam variables,
\begin{eqnarray}
s &=& (p_1 + p_2)^2, \nonumber\\
u &=& (p_1 - p_4 )^2, \nonumber \\
t &=& (p_1-p_3)^2,\nonumber
\end{eqnarray}
subject to the constraint $s+u+t=2(M_{\pi}^2 + M_{\Ka}^2)$. The isospin ${1\over 
2}$ amplitude can be obtained from $T^{3/2}$ by $s \leftrightarrow u$ crossing:
\[
T^{1/2}(s,t,u)={3\over 2} T^{3/2}(u,t,s)-{1\over 2}T^{3/2}(s,t,u).
\]
It is useful to define the amplitudes
\begin{eqnarray}
T^+(s,t,u)&=&{1\over 3}\left( T^{1/2}(s,t,u) + 2 T^{3/2}(s,t,u) \right), \label{nueven} 
\\
T^-(s,t,u)&=&{1\over 3}\left( T^{1/2}(s,t,u) - T^{3/2}(s,t,u) \right), \label{nuodd}
\end{eqnarray}
where $T^+$($T^-$) is even (odd) under $s \leftrightarrow u$ crossing. In the 
$s$-channel the decomposition into partial waves is given by
\begin{equation}\label{partialwaves}
T^I(s(q),t(q,z))=16 \pi \sum_{l=0}^{\infty} (2l+1) t_l^I(q) P_l(z).
\end{equation}
In the expansion of the amplitude to fourth chiral order, only the first three Legendre 
polynomials $P_l$ are needed:
\begin{eqnarray}
P_0(z)&=& 1, \nonumber\\
P_1(z)&=& z, \nonumber \\
P_2(z)&=& {1\over 2}(3 z^2-1).\nonumber
\end{eqnarray}
The dependence of $s$ and $t$ on the momentum $q$ of center of mass and the 
scattering angle $\theta$ is given by
\begin{eqnarray}
s(q)&=& \left( \sqrt{M_{\pi}^2+q^2}+\sqrt{M_{\Ka}^2 +q^2} \right)^2, \nonumber\\
t(q,z)&=& -2q^2 (1-z), \nonumber \\
z&=&\cos \theta.\nonumber
\end{eqnarray} 
In the elastic region, the partial-wave amplitudes satisfy the unitarity 
constraint
\begin{equation}\label{unitarity.const}
{\rm Im}\, t_l^I={2q\over \sqrt{s}} \arrowvert t_l^I \arrowvert,
\end{equation}
and close to the threshold the Taylor coefficients of their real parts 
define the scattering lengths $a$, effective ranges $b$,\dots:
\begin{equation}\label{scatlen.def}
{\rm Re}\, t_l^I(q)= {\sqrt{s(q)} \over 2} q^{2l}\left( a_l^I+b_l^I q^2+c_l^I 
q^4 + O(q^6) \right).
\end{equation}
\subsection{Evaluation of Feynman diagrams}

\begin{figure}
\begin{center}
\epsfig{file=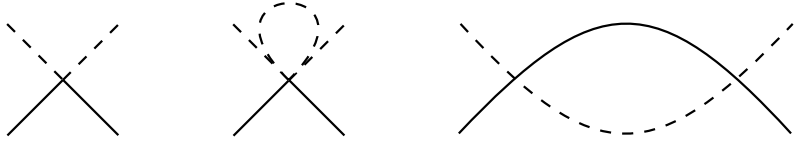,width=.75\textwidth}
\end{center}
\vspace{-5mm}
\caption{{\small {\bf Tree, tadpole and fish diagram.} The continuous and 
dashed lines stand for kaon and pion propagators 
respectively. Crossed fish-diagrams are not shown.}}\label{tretafi.fig}
\end{figure}

For the calculation of the off-shell scattering amplitude it would be necessary 
to evaluate the general form of the matrix elements of two axial currents (or 
pseudoscalar densities) between two kaon states. Here we are only interested in 
the on-shell amplitude and, as explained in \cite{GaLe86}, 
the standard techniques of Feynman diagrams can be used: all external sources are 
turned off in the general Lagrangian (\ref{Lagrangian}) and the traces of the 
matrix fields (\ref{mesonpar}) are developed up to the sixth power in the physical 
meson fields. 
As in the $\chi SU(3)$ theory there are only three different types of Feynman 
diagrams at the one-loop level. These are shown in figure \ref{tretafi.fig}. 
 Tree diagrams must be evaluated using all the vertices of the complete 
fourth order Lagrangian. The tadpole contributions are 
proportional to the pion propagator at $x=0$:
\begin{eqnarray}
{1 \over {\rm i}} \Delta_{\pi}(0) &=& 2 M_{\pi}^2 \lambda + 2 M_{\pi}^2 
{\bar \mu_{\pi}}, \nonumber \\
\lambda &=& {\mu^{d-4} \over 16 \pi^2} \left( {1 \over d-4} - {1\over 2} 
\left( \ln 4\pi + \Gamma'(1) + 1 \right) \right), \nonumber \\
{\bar \mu_{\pi}} &=& {1 \over 32 \pi^2} \ln { M_{\pi}^2 \over \mu^2}. \nonumber
\end{eqnarray}
and their chiral order $D$ is given by $D=D_{\rm tree}+2$ where 
$D_{\rm tree}$ denotes the chiral order of the tree contribution of the same 
vertex. In order to obtain the scattering amplitude to fourth chiral order, 
vertices from ${\cal L}_{\pi K}^{(1)}$ and ${\cal L}_{\pi K}^{(2)}$ must be 
taken into account. Remember that closed kaon 
loops are not calculated in this approach, because such contributions are considered 
to be already included in the coupling constants (see section \ref{kta}).

%
%
%
There are five different fish diagrams contributing to the amplitude according 
to the five different combinations of particles running in the loop. These 
combinations are: 
$\pi^0$--$\pi^0$ ($t$-channel), $\pi^{\pm}$--$\pi^{\mp}$ ($t$-channel), 
$K^0$--$\pi^0$ ($u$-channel), $K^+$--$\pi^+$ ($s$-channel) and 
$K^{\pm}$--$\pi^{\mp}$ ($u$-channel). Figure \ref{tretafi.fig} 
shows the $s$-channel $K^+$--$\pi^+$ diagram which is proportional to

\[
J_{K \pi}(s)={1\over {\rm i}} \int {{\rm d}^d l \over (2 \pi)^d} {1\over M_{\pi}^2 -l^2}{1\over M_{K}^2 - 
(p-l)^2},
\]
where $p=p_1+p_2$ is the total momentum. 

The function $J_{K \pi}(s)$ is Lorentz-invariant and can therefore be evaluated 
at any $p'=\Lambda p$ where $\Lambda$ is an element of the Lorentz group. It is 
expedient to choose the Lorentz transformation such that the initial 
kaon momentum $p_2$ is just $p_2=M_{K} v$, with $v=(1,0,0,0)$, and to develop the 
kaon propagator under the integral sign. The $M_{\Ka}^2$ terms 
in the denominator then cancel and in this 
{\it heavy kaon HK formalism} the function $J_{K \pi}$ becomes
\begin{eqnarray}\label{hubhub}
J_{K \pi}^{HK}(s) \hspace{-2mm}&=&\hspace{-2mm} -{1\over 2 {\rm i} M_{\Ka}} \int {{\rm d}^d l \over (2 \pi)^d} 
{1\over M_{\pi}^2 -l^2}{1\over v(p_1-l)} \left( 1 - {(p_1-l)^2 \over 2 M_{\Ka} v 
(p_1 - l)} + O(l^2) \right). \nonumber \\
&& 
\end{eqnarray}
This integral starts contributing at $O(p^1)$ and the two vertices in the fish 
diagram must therefore be taken from ${\cal L}_{\pi K}^{(m)}$ and 
${\cal L}_{\pi K}^{(n)}$ with $m+n \leq 3$. This confirms the general 
power-counting formula (\ref{chidim}). 
It is shown explicitly in appendix \ref{rel.ap} that the power series resulting 
from the development of the kaon propagator converges to the correct result.
 
The expression (\ref{hubhub}) can be written as a sum of standard integrals of 
heavy-baryon $\chi{\rm PT}$:
\[
J_{K \pi}^{HK}(s) = {1\over 2 M_{\Ka}} J_0(\omega) + {1\over 4 M_{\Ka}^2} (M_{\pi}^2 G_0(\omega) 
-2 \omega G_1(\omega) + d G_2(\omega) + G_3(\omega))
\]
where
\[
\omega = {s-M_{\Ka}^2-M_{\pi}^2 \over 2 M_{\Ka}}={p_1 p_2 \over M_{\Ka}}.
\]
The functions $J_i$ and $G_i$ for $(i = 0, 1, 2)$ are taken from 
\cite{MartinM} and \cite{Meissnerczech} and are given in appendix 
\ref{heavy.app}. 

Posing $\nu = s-u $, which differs from the conventional definition, the full 
fourth order amplitude is given by
{\small
\begin{eqnarray}\label{t32.eqn}
T^{3 \over 2} \hspace{-1mm}&=&\hspace{-1mm} T^{3 \over 2}_1+T^{3 \over 2}_2+T^{3 \over 2}_3+T^{3 \over 
2}_4, \nonumber \\
\nonumber \\
T^{ 3\over 2}_1 \hspace{-1mm}&=&\hspace{-1mm} -{\nu \over 4 F_\pi^2}, \nonumber \\
\nonumber \\
T^{3 \over 2}_2 \hspace{-1mm}&=&\hspace{-1mm} t{A_1 \over 2 F_\pi^2}-\nu^2{{A_2} \over 
16 F_\pi^2} - M_\pi^2 \left( {A_1 \over  F_\pi^2} +{{2 \rho }\over 
F_\pi^2} \right), \nonumber \\
\nonumber \\
T^{3 \over 2}_3 \hspace{-1mm}&=&\hspace{-1mm}  {\nu^2 \over {16 F_\pi^4}} 
\left[ f_s(\nu,t) + 3 f_u(\nu,t) \right] + {\bar J}_{\pi\pi}(t) \left( {{M_\pi^2 \nu}\over 6 F_\pi^2}-{{t \nu}\over 
24 F_\pi^2} \right) + {\bar \mu}_\pi \left( {{t \nu}\over {12 F_\pi^4}} + {\nu^3 \over {16 
F_\pi^4 M_{\Ka}^2 }}\right) \nonumber \\
&&+ \nu^3 \left( {B_2^r\over 32 F_\pi^2} - {1\over {512 \pi^2 F_\pi^4 M_{\Ka}^2}} 
\right) + t \nu \left(  -{B_1^r \over 4 F_\pi^2} + {1\over {384 \pi^2 F_\pi^4 
}}\right) \nonumber \\ 
&&+ M_\pi^2 \nu \left( {B_1^r - 4B_3^r \over 2 F_\pi^2} -{1\over 96 \pi^2 
F_\pi^4 }\right), \nonumber \\
\nonumber \\
T^{3 \over 2}_4 \hspace{-1mm}&=&\hspace{-1mm} f_s(\nu,t) \left[ -{{t \nu}\over {8 F_\pi^4}} + M_\pi^2 \nu 
\left( {1 \over {4 F_\pi^4}} + {\rho \over F_\pi^4} \right) + \nu^3 \left( 
-{1 \over {64 F_\pi^4 M_{\Ka}^2}} + {A_1 \over {32 F_\pi^4 M_{\Ka}^2}}+ {A_2 \over {32 
F_\pi^4}} \right) \right] \nonumber \\
&&+ f_u(\nu,t) \left[ {{3 t \nu}\over {8 F_\pi^4}} + M_\pi^2 \nu 
\left( -{3 \over {4 F_\pi^4}} + {\rho \over F_\pi^4} \right) + \nu^3 \left( 
{3 \over {64 F_\pi^4 M_{\Ka}^2}} + {A_1 \over {32 F_\pi^4 M_{\Ka}^2}}+ {A_2 \over {32 
F_\pi^4}} \right) \right] \nonumber \\
&&+ {\bar J}_{\pi\pi}(t) \left[ t^2 \left( { A_1 \over {2 F_\pi^4}} + { 
{M_{\Ka}^2 A_2} \over {12 F_\pi^4}}\right) + M_\pi t \left( -{{5 
A_1} \over {4 F_\pi^4}} - { 
{3 M_{\Ka}^2 A_2} \over {8 F_\pi^4}} - {{2 \rho } \over F_\pi^4} 
\right)
\right. \nonumber \\
&&+  \left. M_\pi^4 \left( {A_1 \over {2 F_\pi^4}} + {{ M_{\Ka}^2 A_2} \over 
{6 F_\pi^4}} + {\rho \over F_\pi^4}\right) \right] \nonumber \\
&&+ \lambda_{(\nu^4)} \nu^4 + \lambda_{(\nu^2 t)} \nu^2 t + \lambda_{(t^2)} t^2
 + \lambda_{(\nu^2 m^2)} \nu^2 M_{\pi}^2 + \lambda_{(t m^2)} t M_{\pi}^2 +
 \lambda_{(m^4)} M_{\pi}^4, 
\end{eqnarray}}
where
\[
\rho = \sigma_0 \left[ 1-2m_0^2 (C_5^r+2 C_6^r) \right] + O(m_0^4). 
\]

The loop functions are given in terms of the variables $\nu$ and $t$ by
{\small \begin{eqnarray}
f_s(\nu,t) \hspace{-1mm}&=&\hspace{-1mm} \left({1 \over 32 \pi^2} -{\nu \over 64 \pi^2 M_{\Ka}^2}\right) \sqrt{{
(t - \nu )^2 \over M_{\Ka}^4} -16 {M_{\pi}^2 \over M_{\Ka}^2}} \left[ {\rm i}\pi- 
{\rm arccosh}\left( {\nu - t \over 4 M_{\Ka} M_{\pi} }\right)\right], \nonumber \\
\nonumber \\
f_u(\nu,t) \hspace{-1mm}&=&\hspace{-1mm} \left({1 \over 32 \pi^2} +{\nu \over 64 \pi^2 M_{\Ka}^2}\right) \sqrt{{
(t + \nu )^2 \over M_{\Ka}^2}-16 {M_{\pi}^2 \over M_{\Ka}^2} } {\rm arccosh}\left( {
\nu + t \over 4 M_{\Ka} M_{\pi} }\right), \nonumber \\
\nonumber \\
{\bar J}_{\pi \pi}(t) \hspace{-1mm}&=&\hspace{-1mm} {1 \over 16 \pi^2} \left( \sigma(t) \ln {\sigma(t) -1 
\over \sigma(t) +1} + 2 \right), \hspace{7mm} (t \leq 0), \nonumber \\
\sigma(t) \hspace{-1mm}&=&\hspace{-1mm} \sqrt{1-{4 M_{\pi}^2 \over t}}.\nonumber
\end{eqnarray}}

In the physical energy region $\nu + t \geq 4 M_{\Ka} M_{\pi}$ and $t \leq 0$. 
The constants $A_1$, $A_2$, and $\sigma_0$ are scale-independent as are the 
combinations $\lambda_{(pol)}$ of low-energy constants which are 
defined in appendix \ref{leccomb}. It is easy to see whether a given term in 
the scattering amplitude (\ref{t32.eqn}), arises from a tree or a loop diagram: 
loop contributions carry a factor $\pi^2$ in the denominator. As a consistency 
check we have verified that the partial-wave amplitudes $t^I_l$, defined by 
equation (\ref{partialwaves}), satisfy the unitarity constraint 
(\ref{unitarity.const}).

\section{Comparison with the chiral $SU(3)$ amplitude}\label{counting}
A numerical estimate of the low-energy constants can be obtained by 
comparing the above amplitude with the corresponding amplitude of the $\chi 
SU(3)$ theory which is given in \cite{MBK} and whose result is based on the 
generating functional given in \cite{GaLe85}. The standard method of Feynman 
diagrams leads to the same result\footnote{There is a misprint in equation (3.16) 
of \cite{MBK} in the term proportional to $J_{K \eta}^r$ where $3/2$ must be 
replaced by $2/3$ (see also \cite{Borges} page 53).}. The amplitude involves
loop functions of heavy particles ($P, Q \in \{K, \eta \}$),
{\small \begin{eqnarray}
M_{P Q}^r (s) \hspace{-2mm}&=&\hspace{-2mm} {1 \over 12 s}(s-2 \Sigma) {\bar J}_{P Q} (s)+{\Delta^2 \over 3 s^2} 
\; \bar{\hspace*{-1.4mm}\bar{J}}\!_{P Q} (s)-{1 \over 6}k_{P Q} + {1 \over 288 \pi^2}, \nonumber \\
\nonumber \\
L_{P Q} \hspace{-2mm}&=&\hspace{-2mm} { \Delta^2 \over 4 s} {\bar J}_{P Q}, \qquad
K_{P Q} = { \Delta \over 2 s} {\bar J}_{P Q}, \nonumber \\
\nonumber \\
k_{P Q} \hspace{-2mm}&=&\hspace{-2mm} {1 \over 32 \pi^2} { M_P^2 \ln (M_P^2/ \mu^2) - M_Q^2 \ln (M_Q^2/ \mu^2) 
\over M_P^2 - M_Q^2 }, \nonumber \\
\nonumber \\
\; \bar{\hspace*{-1.4mm}\bar{J}}\!_{P Q} (s) \hspace{-2mm}&=&\hspace{-2mm} 
{\bar J}(s) - s {\bar J}'(0), \qquad
{\bar J}'(0) = {1 \over 32 \pi^2} \left( {\Sigma \over \Delta
^2} + 2 { M_P^2 M_Q^2 \over \Delta^3} \ln {M_Q^2 \over M_P^2} \right), \nonumber 
\\
\nonumber \\
{\bar J}_{P Q}(s) \hspace{-2mm}&=&\hspace{-2mm} {1 \over 32 \pi^2} \left( 2 + { \Delta \over s}\ln {M_Q^2 
\over M_P^2 } - {\Sigma \over \Delta} \ln {M_Q^2 \over M_P^2} 
- { \nu \over s} \ln {(s+\nu)^2 - \Delta^2 \over (s-\nu)^2 - \Delta^2} 
\right), \nonumber \\
\nonumber \\
\Sigma \hspace{-2mm}&=&\hspace{-2mm} \Sigma_{P Q} = M_P^2 + M_Q^2,\qquad
\Delta = \Delta_{P Q} = M_P^2 - M_Q^2, \nonumber 
\end{eqnarray}}
\unskip which, close to the threshold, give real
contributions to the scattering amplitude and can be developed in powers of 
$M_{\pi}$, $t$ and $\nu=s-u$. For example, the function ${\bar J}_{\eta,K}$ is 
needed to third $SU(2)$ order:
{\footnotesize \begin{eqnarray}\label{jnk}
&&\hspace{-0.8cm}{\bar J}_{\eta K}(u) = \nonumber \\
&&\hspace{-0.6cm}{ 1 \over \pi^2} \left( {1 \over 16} + 
{\nu \over 32 M_0^2} - {m_0^2 \over 8 M_0^2} + {t \over 32 M_0^2} + {13 \nu^2 \over 1024 M_0^4} - {27 m_0^2 \nu \over 512 M_0^4} + 
{13 t \nu \over 512 M_0^4} + {83 \nu^3 \over 16384 M_0^6} \right) \nonumber \\
&&\hspace{-0.6cm}- {\arctan (\sqrt{2}) \over \sqrt{2} \pi^2} \left(\hspace{-0.3mm} {1 \over 6} 
\hspace{-0.3mm}+\hspace{-0.3mm} {5 \nu \over 96 M_0^2}\hspace{-0.3mm}-\hspace{-0.3mm} 
{11 m_0^2 \over 96 M_0^2} \hspace{-0.3mm}+\hspace{-0.3mm} {5 t \over 96 M_0^2} 
\hspace{-0.3mm}+\hspace{-0.3mm} {53 \nu^2 \over 3072 M_0^4} \hspace{-0.3mm}-\hspace{-0.3mm} 
{35 m_0^2 \nu \over 512 M_0^4} \hspace{-0.3mm}+\hspace{-0.3mm} 
{53 t \nu \over 1536 M_0^4} \hspace{-0.3mm}+\hspace{-0.3mm} 
{343 \nu^3 \over 49152 M_0^6 } \hspace{-0.3mm}\right) \nonumber \\
&&\hspace{-0.6cm}+ {\ln (4/3) \over \pi^2} \left( {5 \over 24} - {\nu \over 192 M_0^2} 
+ {5 m_0^2 \over 24 M_0^2} - {t \over 192 M_0^2} - {\nu^2 \over 384 M_0^4}+{m_0^2 \nu \over 64 M_0^4} - 
{t \nu \over 192 M_0^4} - {\nu^3 \over 768 M_0^6} \right), \nonumber \\
&&
\end{eqnarray}}
\vspace{-1cm}
\begin{eqnarray}
\hspace{-2cm}\mbox{where} \hspace{2cm} M_0^2 &=& B_0 (m_s + {\hat m}), \nonumber \\
m_0^2 &=& 2 B_0 {\hat m}. \nonumber 
\end{eqnarray}

If the two heavy masses are equal, $M_P=M_Q$, the functions ${\bar J}_{P P}$ 
and the constants $k_{P P}$ simplify to
\begin{eqnarray}
{\bar J}_{P P}(t) &=& {t \over 96 \pi^2 M_P^2} + {t^2 \over 960 \pi^2 M_P^4} + 
O(t^3), \nonumber \\
\nonumber \\
k_{P P} &=& {1 \over 32 \pi^2} \left( \ln {M_P^2 \over \mu^2} +1 \right) .\nonumber
\end{eqnarray}
Note that $K$--$K$- and $\eta$--$\eta$-loops can only contribute in the $t$-channel. 

If both amplitudes are developed in powers of $t$, $\nu$ and 
${\hat m}$, an expansion of the $\chi SU(2)$ 
constants in powers of the strange-quark mass is obtained (see appendix \ref{su3estim}).

\section{Threshold parameters}\label{LET}
The formulae (\ref{t32.eqn}) specify the scattering amplitude to fourth chiral 
order. The expression involves 12 combinations of low-energy constants which 
can be related to 12 threshold parameters defined by 
equation (\ref{scatlen.def}). 
These observables are $a^{I}_0$, $b^{I}_0$, $c^{I}_0$, $a^{I}_1$, $b^{I}_1,$ 
and $a^{I}_2$ for $I \in \{1/2, 3/2 \}$.
For $x \in \{a,b,c\}$ and $l \in \{0,1,2\}$ we define
\begin{eqnarray}\label{x+x-}
x^+_l &=& x^{1/2}_l+2 x^{3/2}_l \nonumber \\
x^-_l &=& x^{1/2}_l- x^{3/2}_l, 
\end{eqnarray}
after which it is straightforward to extract these threshold parameters from 
the amplitude. 
\subsection{Corrections to current algebra and a low-energy 
theorem}\label{lep.subsec}
The threshold parameters $x^-_l$ can also be extracted from the amplitude 
$T^-_l (\nu,t)$ which is odd in $\nu$ and which therefore contains the 
polynomials of first and third chiral order ($\nu$, $\nu 
t$, $\nu M_{\pi}^2$, $\nu^3$). Chiral invariance ensures that the lowest-order 
polynomial proportional to $\nu$ is parameter-free, as are the 
leading terms of some of the threshold parameters $x^-_l$. Here only $a_0^-$, 
$a_2^-$ and $b_1^-$ are listed and the parameters $a_1^-$, $b_0^-$ and 
$c_0^-$ are given in appendix \ref{lepar}.
{\small \begin{eqnarray}\label{xmin.lec}
a^-_0 \hspace{-1mm}&=&\hspace{-1mm} {M_{\pi} \over M_{\Ka} + M_{\pi}} 
{3 M_{\Ka} \over 8 \pi F_{\pi}^2} \nonumber \\
&&+ {M_{\pi}^3 \over M_{\Ka} + M_{\pi}} \left( {M_{\Ka} \over 16 \pi^3 F_{\pi}^4}- 
{3 B_1^r M_{\Ka} \over 4 \pi F_{\pi}^2} - {3 B_2^r M_{\Ka}^3 \over 4 \pi 
F_{\pi}^2} + {3 B_3^r M_{\Ka} \over \pi F_{\pi}^2} - {3 M_{\Ka} {\bar \mu}_{\pi} \over 
2 \pi F_{\pi}^4} \right), \nonumber \\
&& \nonumber \\
b^-_1 \hspace{-1mm}&=&\hspace{-1mm} {1 \over M_{\pi} (M_{\Ka} + M_{\pi})} \left(-{1 \over 32 \pi F_{\pi}^2 
M_{\Ka}} -{37 M_{\Ka} \over 11520 \pi^3 F_{\pi}^4} + {B_1^r M_{\Ka} \over 8 \pi F_{\pi}^2} 
- {M_{\Ka} {\bar \mu}_{\pi} \over 24 \pi F_{\pi}^4} \right) \nonumber \\
&&+ {1 \over M_{\Ka} + M_{\pi}} \left( -{85 \over 2304 \pi^3 F_{\pi}^4} 
- {B_1^r \over 8 \pi F_{\pi}^2} - {3 B_2^r M_{\Ka}^2 \over 8 \pi F_{\pi}^2} - 
{17 {\bar \mu}_{\pi} \over 24 \pi F_{\pi}^4} \right),  \\
&& \nonumber\\
a^-_2 \hspace{-1mm}&=&\hspace{-1mm} {1 \over M_{\pi}(M_{\Ka} + M_{\pi})} {M_{\Ka} \over 4800 \pi^3 F_{\pi}^4} + 
{1 \over M_{\Ka} + M_{\pi}} \left( -{1 \over 1152 \pi^3 F_{\pi}^4} + 
{B_1^r \over 20 \pi F_{\pi}^2} - {{\bar \mu}_{\pi} \over 60 \pi F_{\pi}^4} 
\right). \nonumber 
\end{eqnarray}}

It is interesting that loop corrections (factor $\pi^3$ in the denominator) to 
$b_1^-$ have the same chiral order as the tree contribution of lowest order, 
and the $d$-wave scattering length $a_2^-$ is even dominated by a loop 
contribution. This happens because in the definition of these observables (see 
equation (\ref{scatlen.def})) a factor $q^l$ has to be extracted, which in 
general leads to factors of $M_{\pi}$ in the denominator. For the same reason, 
tree contributions from the third-order Lagrangian are less suppressed in 
$b_1^-$ and $a_2^-$ and these observables are therefore the natural candidates 
to replace the unknown coupling constants $B_1^r$ and $B_2^r$ in the expressions 
for the four other observables. The remaining 
low-energy constant $B_3^r$ corresponds to a chiral-symmetry breaking term and 
therefore cannot be extracted from the angular and energy distributions of the 
scattering cross section. However, the theoretical estimate from 
appendix \ref{su3estim},
\[ 
B_3^r(\mu=770) = 7.9 \cdot 10^{-7} \hspace{1mm}{\rm MeV}^{-2},
\]
gives parameterless corrections to the `current algebra' result of the 
observables $a_0^-$, $b_0^-$, $a_1^-$ and $c_0^-$. Here only $a_0^-$ is listed. 
The parameters $a_1^-$ and $b_0^-$ are given in appendix \ref{lepar}.
{\small \begin{eqnarray}\label{aba}
a^-_0 \hspace{-2mm}&=&\hspace{-2mm} {M_{\pi} \over M_{\Ka} + M_{\pi}} \left({3 M_{\Ka} \over 8 \pi F_{\pi}^2} + 
{M_{\Ka}^3 \over 960 \pi^3 F_{\pi}^4} \right) + {M_{\pi}^2 \over M_{\Ka} + M_{\pi}} \left({1 \over 16 \pi F_{\pi}^2} + 
{M_{\Ka}^2 \over 240 \pi^3 F_{\pi}^4} -5 M_{\Ka}^3 a_2^- \right) \nonumber \\
&&+ {M_{\pi}^3 \over M_{\Ka} + M_{\pi}} \left( {3 B^r_3 M_{\Ka} \over \pi 
F_{\pi}^2} + {49 M_{\Ka} \over 384 \pi^3 F_{\pi}^4}- {M_{\Ka} {\bar \mu}_{\pi} \over 4 \pi F_{\pi}^4} - 15 M_{\Ka}^2 a_2^- + 2 
M_{\Ka}^2 b_1^- \right) \nonumber \\
&&+ {M_{\pi}^4 \over M_{\Ka} + M_{\pi}} \left( -10 M_{\Ka} a_2^- + 2 M_{\Ka} b_1^- \right). 
\end{eqnarray}}
\unskip The only input from $\chi SU(3)$ is the constant $B^r_3$, which appears 
in the term of $O(p^3)$. 

Another interesting combination between scattering lengths is 
\[
A^-=a_0^- -6 M_{\pi} M_{\Ka} a_1^- + 30 M_{\pi}^2 M_{\Ka}^2 a_2^- ,
\]
which is discussed (in the $\chi SU(3)$ current algebra (CA) approach) by Lang 
\cite{Lang}. Without taking into account loop corrections
\[
 A^-_{CA} =0.
\]
From the results (\ref{xmin.lec}) and (\ref{xmin.lec2}) one finds 
{\small\begin{eqnarray}
A^- \hspace{-1mm}&=&\hspace{-1mm} {M_{\pi} \over M_{\Ka} + M_{\pi}} {M_{\Ka}^3 \over 160 \pi^3 F_{\pi}^4} 
+ {M_{\pi}^3 \over M_{\Ka} + M_{\pi}} \left( {3 M_{\Ka} \over 32 \pi^3 F_{\pi}^4} 
+ {3 B_2^r M_{\Ka}^3 \over 2 \pi F_{\pi}^2} + {3 M_{\Ka} {\bar \mu}_{\pi} \over \pi 
F_{\pi}^4}\right),\nonumber
\end{eqnarray}}
\unskip valid to fourth chiral order.
The chiral-symmetry breaking terms (proportional to $B_3^r$) cancel in this 
sum. 
The first- and the third-order loop corrections are $0.009 M_{\pi}^{-1}$ and 
$0.011 M_{\pi}^{-1}$ respectively. The correction from the polynomial $\nu^3$ -- which is of 
$\chi SU(3)$ order $p^6$ -- can be estimated with the $\chi SU(3)$ value for 
$B_2^r$ and is about 10 \% of the loop corrections. This leads to 
\[
A^- = 0.022 M_{\pi}^{-1}.
\]

Alternatively, the unknown constant $B_2^r$ can be expressed in terms of the 
observables $a_2^-$ and $b_1^-$,
{\small \begin{eqnarray}\label{exact}
A^- \hspace{-1mm}&=&\hspace{-1mm} {M_{\pi} \over M_{\Ka} + 
M_{\pi}} {M_{\Ka}^3 \over 240 \pi^3 F_{\pi}^4} + {M_{\pi}^2 \over M_{\Ka} + M_{\pi}} 
\left( -{1 \over 8 \pi F_{\pi}^2} + 10 
M_{\Ka}^3 a_2^- -{M_{\Ka}^2 \over 480 \pi^3 F_{\pi}^4} \right) \nonumber \\
&&- {M_{\pi}^3 M_{\Ka} \over M_{\Ka} + M_{\pi}} \left( 4 M_{\Ka} b_1^- + 
{ 1 \over 16 \pi^3 F_{\pi}^4} \right)- {M_{\pi}^4 M_{\Ka} \over M_{\Ka} + M_{\pi}} \left( 
10 a_2^- + 4 b_1^- \right),
\end{eqnarray}}
\unskip giving a result which is exact to fourth $\chi SU(2)$ order and which 
involves no low-energy constants.

Table \ref{threshpar.min} shows the numerical values obtained by using 
$\chi SU(3)$ estimates for $B_1^r$, $B_2^r$ and $B_3^r$ in 
(\ref{xmin.lec}) and (\ref{xmin.lec2}), given in the appendix. 
Some experimental values for $a_0^-$ are given 
in table \ref{thresh.exp}. More accurate data on threshold parameters of higher 
order would be needed to test relations (\ref{aba}), (\ref{exact}) and 
(\ref{aba2}).

\begin{table}[!]
\caption{\footnotesize {\bf Threshold parameters associated with the amplitude $T^-$.} 
There are tree ({\rm CA}) and loop contributions (loops) from 
${\cal L}_{\pi \Ka}^{(1)}$. The total result ($\chi SU(2)+\chi SU(3)$) also 
contains polynomial contributions from ${\cal L}_{\pi \Ka}^{(3)}$, with 
low-energy constants estimated from $\chi SU(3)$ at a renormalisation scale of 
$\mu=770 {\rm MeV}$. }\label{threshpar.min}
\begin{center}  \begin{tabular}{|c|c|c|c|} \hline
\hspace{1mm} thresh. param. \hspace{1mm} & \hspace{7mm} CA \hspace{7mm} & 
\hspace{7mm}loops \hspace{3mm}      &\hspace{6mm} $\chi SU(2) + \chi SU(3)$ 
\hspace{6mm} \\ \hline \hline
$a_0^- M_{\pi}$    & 0.20  & 0.01        &$0.23 \pm 0.01$     \\  \hline
$b_0^- M_{\pi}^{3}$& 0.10  & 0.01        &$0.10 \pm 0.01$    \\ \hline
$c_0^- M_{\pi}^{5}$& -0.03 & -0.01       &$-0.052 \pm 0.003$  \\ \hline
$a_1^- M_{\pi}^{3}$& 0.009 & -0.001      &$0.013 \pm 0.002$   \\ \hline
$b_1^- M_{\pi}^{5}$&-0.001 & -0.001      &$-0.002 \pm 0.001$  \\ \hline
$a_2^- M_{\pi}^{5}$&   0   &$2 \cdot 10^{-5}$ & $(20 \pm 7) \cdot 10^{-5}$ \\ \hline
\end{tabular} \end{center}
\end{table}

The values for the scattering lengths are consistent with the result in 
\cite{BKMthresh}, where the slope parameters $b_0^{1/2}$ and $b_0^{3/2}$, 
however, were obtained as 
\begin{equation}\label{bmbk}
(b_0^{1/2})_{\scriptsize \cite{BKMthresh}} = (0.14 \pm 0.02) M_{\pi}^{-3}, 
\mbox{ } \quad (b_0^{3/2})_{\scriptsize \cite{BKMthresh}} = (-0.011 \pm 0.005) 
M_{\pi}^{-3}, 
\end{equation}
which is inconsistent with our result. In the $\chi SU(3)$ case 
$M_{\pi}/M_{\Ka}$ is not a chiral expansion parameter and 
the analytical expressions for the threshold parameters are therefore more 
voluminous then in the $\chi SU(2)$ case. Thus only numerical values are 
given in \cite{BKMthresh}.

\begin{table}[h]
\caption{\footnotesize {\bf The scattering length $a_0^-$ as determined in various analyses.} 
The values are in chronological order. The interval in the second column is chosen 
so as to cover the experimental results for $a^{(1/2)}_0$ \cite{Bingham} and 
$a^{(3/2)}_0$ \cite{Bakker}. References 
\cite{Karabarbounis} and \cite{Krivoruchenko} are dispersion relation analyses 
based on more recent experiments, \cite{Estabrooks} and \cite{Aston, 
Estabrooks} respectively, with much more data. The error in the second column 
is chosen so as to cover the results for $a^{(1/2)}_0$ and 
$a^{(3/2)}_0$ given in \cite{Karabarbounis}.}\label{thresh.exp}
\begin{center} \begin{tabular}{|c|c|c|c|} \hline
thresh. param.  &  before 1977 & \hspace{0.5mm}Karabarbounis 
\cite{Karabarbounis}\hspace{0.5mm} &\hspace{1mm} Krivoruchenko \cite{Krivoruchenko} 
\hspace{1mm} \\ \hline \hline
$a_0^- M_{\pi}$ &$0.2 \dots 0.4$& $0.26 \pm 0.06$                   &$0.22 \pm 0.02$ \\  \hline
\end{tabular} \end{center}
\end{table}

\subsection{Other parameter-free relations between threshold parameters}
The amplitude $T^+_l$ is even in $\nu$ and it therefore involves the 
polynomials of second and fourth chiral order. In this case, chiral invariance 
does not restrict the associated threshold parameters, which are listed in 
appendix \ref{lepar}.

The quantity $A^+$, defined by
{\small \begin{eqnarray}
A^+ \hspace{-2mm}&=&\hspace{-2mm} a_0^+ - M_{\pi}^2 b_0^+ + {3 M_{\pi}^3 \over 2 M_{\Ka}} \left( a_1^+ + 
{b_0^+ \over 4} \right) \nonumber \\ 
&&\hspace{-2mm}+ M_{\pi}^4 \left[ c_0^+ - 10 a_2^+ 
+ \left( {1 \over 288 \pi^2 F_{\pi}^2} - { 5 
\over 16 M_{\Ka}^2} \right) b_0^+ - \left( {1 \over 48 \pi^2 F_{\pi}^2} + 
{ 9 \over 8 M_{\Ka}^2} \right) a_1^+ \right], \nonumber
\end{eqnarray}}
\unskip preserves chiral-power 
counting in the sense that no threshold parameter is needed to an unnaturally 
high precision. This quantity contains only (combinations of) coupling constants 
associated with chiral-symmetry breaking terms:   
{\small \begin{eqnarray}\label{A+}
A^+ \hspace{-3mm}&=&\hspace{-2mm} -{M_{\pi}^2 \over M_{\Ka} + M_{\pi}} {3 \sigma_0 \over 4 \pi F_{\pi}^2} - 
{M_{\pi}^3 \over M_{\Ka} + M_{\pi}} {15 \sigma_0 \over 32 \pi F_{\pi}^2 M_{\Ka}} 
\nonumber \\
&&\hspace{-2mm}+ {M_{\pi}^4 \over M_{\Ka} + M_{\pi}} \left( {3 \sigma_0 \over 64 \pi F_{\pi}^2 
M_{\Ka}^2} + {3 \sigma_0 (C_5^r + 2 C_6^r) \over 2 \pi F_{\pi}^2 } + { 3 
\lambda_{(t m^2)} \over 4 \pi} + { 3 \lambda_{(m^4)} \over 8 \pi} + 
{ \sigma_0 \over 128 \pi^3 F_{\pi}^4 } \right) \nonumber \\
\hspace{-3mm}&=&\hspace{-2mm} (0.03 \pm 0.02) M_{\pi}^{-1} + (0.006 \pm 0.004) M_{\pi}^{-1} 
+ (0.004 \pm 0.006) M_{\pi}^{-1} \nonumber \\
\hspace{-3mm}&=&\hspace{-2mm} (0.04 \pm 0.03) M_{\pi}^{-1}.
\end{eqnarray}}
\unskip The numerical values again result from theoretical $\chi SU(3)$ estimations of the 
$\chi SU(2)$ constants.

The same estimations can be used to obtain numerical values for the threshold 
parameters from expressions (\ref{ahoch+}). However, due to the absence of 
parameter-free lowest-order predictions these values, shown in table 
\ref{threshpar.plu}, are less precise than 
those discussed in subsection \ref{lep.subsec}. 

The parameter 
\[a_1^{(1/2)}=(a_1^+ + 2 a_1^-)/3=(0.018 \pm 0.002 )M_{\pi}^{-3}
\]
is in very good agreement with the dispersion relation 
analysis of Karabarbounis and Shaw \cite{Karabarbounis}:

\[ 0.0178 M_{\pi}^{-3} \leq  (a_1^{(1/2)})_{\scriptsize \cite{Karabarbounis}} 
\leq 0.0185 M_{\pi}^{-3}. \]

Mean experimental values for $a_0^{1/2}$ and $a_0^{3/2}$ are given in 
\cite{MBK}:

\[ a_0^{1/2}=0.13 \dots 0.24 M_{\pi}^{-1},\quad a_0^{3/2}=-0.13 \dots -0.05 
M_{\pi}^{-1}. \]

These were obtained long before the high-statistics experiment of \cite{Aston} in 
1988 which is an important reference for $\pi$--$K$ resonance data collected by the PDG 
\cite{PDG}. The absence of precise experimental data for higher-order threshold 
parameters precludes the verification of relation (\ref{A+}).

\begin{table}[!]
\caption{\footnotesize {\bf Threshold parameters associated with the amplitude $T^+$.} 
Current algebra gives no contributions and loops must be considered form 
${\cal L}_{\pi \Ka}^{(1)}$ {\it and} ${\cal L}_{\pi \Ka}^{(2)}$. The total result 
($\chi SU(2)+\chi SU(3)$) contains in addition polynomial contributions from 
${\cal L}_{\pi \Ka}^{(2)}$ and ${\cal L}_{\pi \Ka}^{(4)}$, with low-energy constants 
estimated from $\chi SU(3)$
. }\label{threshpar.plu}
\begin{center}  \begin{tabular}{|c|c|c|c|} \hline
thresh. param.  &\hspace{6mm} $\chi SU(2)+\chi SU(3)$ \hspace{6mm} \\ \hline \hline
$a_0^+ M_{\pi}$    & $0.08 \pm 0.04$     \\  \hline
$b_0^+ M_{\pi}^{3}$& $0.06 \pm 0.03$    \\ \hline
$c_0^+ M_{\pi}^{5}$& $0.015 \pm 0.004$  \\ \hline
$a_1^+ M_{\pi}^{3}$& $0.023 \pm 0.008$   \\ \hline
$b_1^+ M_{\pi}^{5}$& $0.003 \pm 0.002$  \\ \hline
$a_2^+ M_{\pi}^{5}$& $(4 \pm 3) \cdot 10^{-4}$ \\ \hline
\end{tabular} \end{center}
\end{table}

As in the case of $T_{\pi K}^-$, the scattering lengths are consistent with 
the result of \cite{BKMthresh}, whereas, as can be seen from equation 
(\ref{bmbk}) and definition (\ref{x+x-}), the effective range parameter $b_0^+$ 
is half the size.
This cannot be due to the fact that the authors of 
\cite{BKMthresh} use older values for the $\chi SU(3)$ constants, since the effect of this 
would be too small. In the $\chi SU(3)$ case, our calculation of the 
$T_{\pi K}^{3/2}$ amplitude coincides exactly with the one in \cite{MBK}.

\section{Summary and Conclusion}\label{summary}
To the best of our knowledge, this is the first analysis of the threshold 
structure of pion-kaon scattering using only $\chi SU(2)$ symmetry 
of the strong interactions. Chiral $SU(2)$ symmetry is less strongly broken in 
nature than $\chi SU(3)$ symmetry and the chiral expansion is therefore expected to 
converge more rapidly. Chiral $SU(2)$ symmetry only allows for one interaction 
term of $O(p^1)$,
\[
{\cal L}^{(1)}_{\pi,K} = D_{\mu} K^+ D^{\mu} K - M^2 K^+ K,
\]
where $D_{\mu} K= \partial_{\mu} K + \Gamma_{\mu} K$ is the covariant derivative 
containing the pion field (definition \ref{Gamma}).

The lowest-order contribution to the isospin $3/2$ scattering amplitude is 
therefore parameter-free:
\[
T^{3 \over 2}_1 = -{\nu \over 4 F_\pi^2 },
\]
where $\nu=s-u$ contains indeed a term proportional to the pion mass and 
is therefore of first chiral order\footnote{In the standard convention the variable $\nu$ is defined by 
$\nu =(s-u)/4 M_{\Ka}$.}. The lowest-order contributions to the threshold 
parameters do not differ significantly from those given by the $\chi SU(3)$ 
current algebra result \cite{Lang}. 

The isospin $3/2$ amplitude was calculated to fourth chiral order in the 
isospin limit and without including elecromagnetic interactions. The result is 
given by equation (\ref{t32.eqn}).

The loop correction to the current algebra result for the scattering length 
$a_0^-$ amounts to $5\%$ at a renormalisation scale of 
$\mu=770{\rm MeV}$. The loop contributions to 
the parameters $b_0^-$, 
$c_0^-$, $a_1^-$, $b_1^-$ and $a_2^-$ are given in table \ref{threshpar.min}.
The exact result for theses observables -- valid to fourth chiral order -- 
involves three unknown low-energy constants, $B_1^r$, $B_2^r$ and $B_3^r$. 
The observables $b^-_1$ and $a^-_2$ measure the first two constants so that a 
theoretical estimate (from chiral $SU(3)$ symmetry) of the ``symmetry-breaking 
constant" $B_3^r$ produces parameter-free predictions for $a^-_0$, $b^-_0$ and 
$a^-_1$. These relations are given by formulae (\ref{aba}) and (\ref{aba2}). 
There is also a pure 
$\chi SU(2)$ low-energy theorem -- given by formula (\ref{exact}) -- 
involving the above observables.
 
The $\nu$-even amplitude $T_{\pi K}^+$ starts at second chiral order and the associated 
threshold parameters $a^+_0$, $b^+_0$, $c^+_0$, $a^+_1$, $b^+_1$ and $a^+_2$ 
are not restricted by chiral $SU(2)$ symmetry. However, there are relations between 
these observables involving only $\chi SU(2)$ constants associated with a 
chiral-symmetry-breaking term. Equation (\ref{A+}) shows the result when all these 
constants are estimated from $\chi SU(3)$.

The relations between threshold parameters (\ref{aba}, \ref{exact}, \ref{A+} 
and \ref{aba2}), valid to fourth 
chiral ($SU(2)$) order, are the main result of the present analysis and could 
not have been obtained within $\chi SU(3)$. In relations (\ref{aba}), 
(\ref{A+}) and (\ref{aba2}) a theoretical $\chi SU(3)$ input 
is used (only for $\chi SU(2)$ symmetry-breaking terms) whereas 
relation (\ref{exact}) is completely free of any $\chi SU(3)$ approximation.

The present phenomenological data for the threshold parameters do not 
allow for a test of the obtained relations between threshold parameters.
An analysis based on dispersion relations, such as was done for example 
in 1978 by Johannesson and Nilsson \cite{Johannesson2} but including all 
recent data, might improve this situation \cite{MBK}.

Numerical estimates for our low-energy constants were obtained by 
matching the scattering amplitude with the corresponding {$\chi SU(3)$} result. 
The numerical results for {\it individual} threshold parameters, where these 
estimates are used, should therefore be compatible with the 
{$\chi SU(3)$} result. This is the case for all scattering lengths. There is 
however an incompatibility between our results for the effective ranges and 
those given in \cite{BKMthresh}. In that reference no {\it analytic} 
expressions are given for the effective ranges.

Another possibility to obtain estimates for the low-energy constants 
in the context of chiral $SU(2)$ perturbation theory 
would be to use a {\it resonance 
saturation} approach, similar to the one performed by Bernard and Kaiser 
\cite{BerKai} in the $\chi SU(3)$ case. 

I would like to thank Heinrich Leutwyler, who suggested the subject of this 
work, for his continual help as well as Gerhard Ecker and G\'{e}rard Wanders 
for useful discussions and stimulating comments.

\appendix
\section{Non-relativistic terms}\label{nonrelstr}
Non-relativistic terms in the Lagrangian have the advantage that they correspond to a well 
defined chiral order. Lorentz invariance can be recovered at a later stage by 
imposing the necessary relations on the coupling constants. This is done 
implicitly in subsection \ref{Lagdens4.sub} by writing the Lagrangian in 
manifest relavistically invariant form. In the following, all allowed terms 
relevant for the present application and respecting chiral, parity and 
charge-conjugation invariance are shown, classified by their chiral order. A 
covariant derivative counts as $O(p^1)$ and the field $\chi$ indicating 
chiral-symmetry breaking terms as $O(p^2)$. The definitions of the 
fields involving $k$ are given by the corresponding relations in table 
\ref{blocks} with $k$ substituted for $K$. Terms of second or higher chiral 
order, proportional to $v_\mu k^\mu$ can be eliminated by appropriate 
redefinitions of the field $k$. Such equation-of-motion terms are not listed in the 
following.

\begin{itemize}
\item{Order $p^0$:} \hspace{1cm} $k^+k$
\item{Order $p^1$:} \hspace{1cm} $v_{\mu}\bra k_-^{\mu}\ket$
\item{Order $p^2$:} 
\begin{equation}\label{nunu}
\left.
\begin{array}{ll}
\bra k_{+,\mu}^\mu \ket,& \\
\bra \Delta_{\mu} \Delta^{\mu} \ket k^+k, & v^\mu v^\nu \bra \Delta_{\mu} 
\Delta_{\nu} \ket k^+k, \\
\bra \chi_+ kk^+\ket, & \bra \chi_+ \ket k^+k 
\end{array} \right\}. 
\end{equation}
Lorentz invariance implies that the first term in (\ref{nunu}) has the 
same coupling constant as the term of first order, 
the second line contains the two 
chirally symmetric terms generating the polynomials $t$ and $\nu^2=(s-u)^2$ in 
the scattering amplitude; the last line generates the chiral-symmetry 
breaking terms i.e. terms proportional to ${\hat m}$. In our case, where 
$m_u=m_d$, the two terms are in fact proportional. Our final 
Lagrange density is therefore more general than it need be for the 
present application.
\item{Order $p^3$:}
\begin{equation}\label{jubjub}
\left.
\begin{array}{ll}
v^\mu \bra \Delta_\mu \Delta_\nu \ket \bra k_-^\nu \ket, & \\
v^\mu \bra [ \Delta_{\mu \nu}, \Delta^\nu ] kk^+ \ket, & v^\mu v^\nu v^\rho 
\bra [ \Delta_{\mu \nu},\Delta_\rho ] kk^+ \ket, \\
v^\mu \bra [ \chi_-,\Delta_\mu ] kk^+ \ket &
\end{array} \right\}.
\end{equation}
The first term in (\ref{jubjub}) is related by Lorentz invariance to a 
second-order term and the last three terms correspond to the three 
polynomials $t \nu$, $\nu^3$ and $M_{\pi}^2 \nu$.
\end{itemize}
\subsection{Order $p^4$}
\subsubsection{Terms related by relativistic invariance}
\[
\begin{array}{lll}
\bra \Delta_\mu \Delta_\nu \ket \bra k_+^{\mu \nu} \ket,&& \\
\bra [ \Delta_{\mu \nu}, \Delta^\nu ] k_-^\mu \ket,&v^\mu v^\nu \bra 
\Delta_{\mu \nu} \Delta_\rho \ket \bra k_+^\rho \ket,&\bra [\chi_-
,\Delta_\mu] k_-^\mu \ket, \\
v^\mu v^\nu \bra \chi_- \Delta_{\mu \nu} \ket k^+ k.&&
\end{array}
\]
The first, second and third lines are related to terms of order two, three 
and four respectively. 
\subsubsection{Chirally symmetric terms corresponding to the polynomials 
$t^2$, $t \nu^2$ and $\nu^4$}
\begin{eqnarray}
\bra \Delta_{\mu \nu} 
\Delta^\nu \ket \bra k_+^\mu \ket,& v^\mu v^\nu \bra 
\Delta_{\mu \rho} \Delta_\nu \ket \bra k_+^\rho \ket, & v^\mu v^\nu v^\rho 
v^\sigma \bra \Delta_{\mu \nu} \Delta_{\rho \sigma} \ket k^+ k \nonumber
\end{eqnarray}
\subsubsection{Chiral-symmetry breaking terms with one $\chi$ generating 
$M_{\pi}^2 t$ and $M_{\pi}^2 \nu^2$}
\begin{eqnarray}
&& \begin{array}{lll}
\bra \chi_+ k_{+\mu}^\mu \ket,& \bra \chi_+ \ket \bra 
k_{+\mu}^\mu \ket,&\bra \chi_- \Delta_\mu \ket \bra 
k_+^\mu \ket, \\
\bra \Delta_\mu \Delta^\mu \ket \bra \chi_+ k k^+ \ket,& \bra \Delta_\mu 
\Delta^\mu \ket \bra \chi_+ \ket k^+ k,&v^\mu v^\nu \bra \hspace{-0.5mm}
\Delta_\mu \Delta_\nu \ket \bra \chi_+ k k^+ \hspace{-0.5mm}\ket,\\
v^\mu v^\nu \bra \Delta_\mu 
\Delta_\nu \ket \bra \chi_+ \ket k^+ k,&
v^\mu v^\nu \bra\hspace{-1mm} \{\hspace{-0.5mm} \Delta_\mu \Delta_\nu \} 
\{ \chi_+ k k^+ \hspace{-0.5mm}\}\hspace{-1mm} \ket & \\
\end{array} \nonumber
\end{eqnarray}
\subsubsection{Chiral-symmetry breaking terms with two $\chi$ generating 
$M_{\pi}^4$}
\begin{eqnarray}
&& \begin{array}{llll}
\bra \chi_+\ket k^+ \chi_+ k,& \bra \chi_+^2 \ket k^+ k,& \bra \chi_+ \ket^2 
k^+ k,& \bra \chi_-^2 \ket k^+ k
\end{array} \nonumber
\end{eqnarray}
\section{Heavy-baryon loop integrals}\label{heavy.app}
For applications to pion-kaon scattering, only a small number of one-loop integrals 
used in heavy baryon $\chi{\rm PT}$ is needed. This is due to the absence of cubic 
pion-kaon couplings. The relevant integrals can be found in \cite{MartinM} and 
\cite{Meissnerczech}.
\begin{eqnarray}
{1\over {\rm i}} \int {{\rm d}^d l \over (2 \pi)^d} { \{ 1,l_{\mu},l_{\mu}, 
l_{\nu} \}\over (M_{\pi}^2-l^2)(v \cdot l-\omega)} &=& \{ J_0(\omega),v_{\mu} 
J_1(\omega), g_{\mu \nu} J_2(\omega) + v_{\mu} v_{\nu} J_3(\omega) \} 
\nonumber \\
{1\over {\rm i}} \int {{\rm d}^d l \over (2 \pi)^d} { \{ 1,l_{\mu},l_{\mu}, 
l_{\nu} \}\over (M_{\pi}^2-l^2)(v \cdot l-\omega)^2} &=& \{ G_0(\omega),v_{\mu} 
G_1(\omega), g_{\mu \nu} G_2(\omega) + v_{\mu} v_{\nu} G_3(\omega) \}\nonumber 
\end{eqnarray} 

{\small \begin{eqnarray}
J_0(\omega)\hspace{-2mm}&=&\hspace{-2mm} -4 \omega \lambda + {\omega \over 8 \pi^2} \left( 1-\ln {M_{\pi}^2 
\over \mu^2}\right) + {\sqrt{\omega^2-M_{\pi}^2 } \over 4 \pi^2} {\rm arccosh}{-\omega 
\over M_{\pi}} ,\hspace{15.6mm} [ \omega < -M_{\pi} ], \nonumber \\
&=&\hspace{-2mm} -4 \omega \lambda + {\omega \over 8 \pi^2} \left( 1-\ln {M_{\pi}^2 \over 
\mu^2}\right) - {\sqrt{M_{\pi}^2 - \omega^2} \over 4 \pi^2} 
{\rm arccos}{-\omega 
\over M_{\pi}} ,\hspace{17.7mm} [ \omega^2 < M_{\pi}^2 ], \nonumber \\
&=&\hspace{-2mm} -4 \omega \lambda + {\omega \over 8 \pi^2}  
\left( 1-\ln {M_{\pi}^2 \over \mu^2}\right) - 
{\sqrt{\omega^2-M_{\pi}^2 } \over 4 \pi^2} \left({\rm arccosh}{\omega 
\over M_{\pi}} - {\rm i} \pi \right),\hspace{1mm} [\omega < -M_{\pi}], \nonumber \\
J_1(\omega)\hspace{-2mm}&=&\hspace{-2mm} \omega J_0(\omega) + {1\over {\rm i}} \Delta_{\pi} (0), 
\nonumber \\
J_2(\omega)\hspace{-2mm}&=&\hspace{-2mm} {1\over d-1}\left[ (M_{\pi}^2 - 
\omega^2 )J_0(\omega) - \omega {1\over {\rm i}} 
\Delta_{\pi} (0) \right], \nonumber \\
J_3(\omega)\hspace{-2mm}&=&\hspace{-2mm} \omega J_1(\omega) - J_2(\omega),\nonumber 
\end{eqnarray}}
where
\begin{eqnarray}
{1\over {\rm i}} \Delta_{\pi} (0) &=& 2 M_{\pi}^2 (\lambda + {1\over 32 \pi^2} 
\ln({M_{\pi}^2 \over \mu^2})) \nonumber \\
\lambda &=& {\mu^{d-4}\over 16 \pi^2} \left( {1 \over d-4} - {1\over 2}( 
\Gamma '(1) + 1 + \ln(4 \pi)) \right).\nonumber
\end{eqnarray}

The functions $G_i (i=0,1,2,3)$ can be obtained using the identity
\begin{eqnarray}
{1 \over (v \cdot l - \omega)^2} &=& {\partial \over \partial \omega} \left( {1 
\over v \cdot l - \omega} \right); \nonumber \\
G_i(\omega)&=& {\partial J_i \over \partial \omega }(\omega).\nonumber
\end{eqnarray}

\section{Combinations of low-energy constants contributing at fourth chiral 
order}\label{leccomb}
The constants $\lambda_{(pol)}$ contributing at fourth order to the scattering 
amplitude given by equation (\ref{t32.eqn}) are defined by the 
combinations of coupling constants
{\footnotesize \begin{eqnarray}
\lambda_{(\nu^4)} \hspace{-2mm}&=&\hspace{-2mm} -{1 \over {1024 \pi^2 F_\pi^4 
M_{\Ka}^4}} - {C_4^r \over {128 F_\pi^2}} + {\bar \mu}_\pi {3 \over 
{32 F_\pi^4 M_{\Ka}^4}}, \nonumber \\
\lambda_{(\nu^2 t)} \hspace{-2mm}&=&\hspace{-2mm} -{3 \over {256 \pi^2 F_\pi^4 M_{\Ka}^2}} 
+ {C_2^r \over {32 F_\pi^2}}- { C_3^r \over {32 F_\pi^2}}  + {\bar \mu}_\pi {{3}\over {8 F_\pi^4 M_{\Ka}^2}}, \nonumber \\
\lambda_{(t^2)} \hspace{-2mm}&=&\hspace{-2mm} -{A_1 \over {32 \pi^2 F_\pi^4}} 
-{A_2 \over {16 F_\pi^2}} - {{ M_{\Ka}^2 A_2} \over {192 \pi^2 F_\pi^4}} + 
{C_1^r \over {4 F_\pi^4}} + {\bar \mu}_\pi \left(-{A_1 \over F_\pi^4} -
{{M_{\Ka}^2 A_2} \over {6 F_\pi^4}} \right) \nonumber \\
\lambda_{(\nu^2 m^2)}\hspace{-2mm}&=&\hspace{-2mm} -{{3 (B_2^r+C_3^r)} 
\over {32 F_\pi^2}} - {{C_{10}^r +2C_{11}^r+2C_{12}^r }\over {8 F_\pi^2}} + 
{{A_2 (C_5^r +2C_6^r)} \over {8 F_\pi^2}} + {\bar \mu}_\pi 
\left({A_2 \over {4 F_\pi^4}}-{5\over{4 F_\pi^4 M_{\Ka}^2}} \right), \nonumber \\
\lambda_{(t m^2)} \hspace{-2mm}&=&\hspace{-2mm} - {C_1^r \over {2 F_\pi^2}} + 
{{2 C_7^r}\over F_\pi^2} + {C_8^r \over F_\pi^2} + {{2 C_9^r}\over 
F_\pi^2} + {{C_5^r+2C_6^r}\over F_\pi^2} - {{A_1 
(C_5^r + 2 C_6^r) } \over F_\pi^2} \nonumber \\
&&\hspace{3mm}+{{5 A_1} \over {64 \pi^2 F_\pi^4}} + {{3 M_{\Ka}^2 
A_2}\over {128 \pi^2 F_\pi^4}} + {\rho \over {8 \pi^2 
F_\pi^4}} +{\bar \mu}_\pi \left(-{1\over F_\pi^4} + {{5 A_1} \over {2 
F_\pi^2}} + {{13 M_{\Ka}^2 A_2} \over {12 F_\pi^4}} + {{4 \rho } 
\over F_\pi^4} \right), \nonumber \\
\lambda_{(m^4)} \hspace{-2mm}&=&\hspace{-2mm}  - 
{{2 (C_8^r+2C_9^r)} \over F_\pi^2} - {{16 (C_{13}^r+C_{14}^r+2C_{15}^r+C_{16}^r)} 
\over F_\pi^2} + 2{{(A_1+2\rho )(C_5^r+2C_6^r)} 
\over F_\pi^2} + {{4 \rho l_3^r}\over F_\pi^4} \nonumber \\
&&\hspace{3mm}-{A_1 \over {32 \pi^2 F_\pi^4}} -{{M_{\Ka}^2 A_2} 
\over {96 \pi^2 F_\pi^4}} -{\rho \over {16 \pi^2 F_\pi^4}}+ {\bar \mu}_\pi \left({2\over F_\pi^4} + {{2 A_1} \over 
F_\pi^2} - {{M_{\Ka}^2 A_2} \over {2 F_\pi^4}} + {{6 \rho } \over 
F_\pi^4} \right). \nonumber 
\end{eqnarray}}
\unskip The scale 
dependence of the low-energy constants can easily be seen, since 
all the above combinations are scale independent, as are $A_1$, $A_2$ 
and $\sigma_0$.

\section{Estimates for the low-energy constants from $\chi 
SU(3)$}\label{su3estim}
The development of (combinations of) the $\chi SU(2)$ constants associated with 
the Lagrangian ${\cal L}_{\pi K}$ in powers of the strange-quark mass, as obtained 
from the comparison of the scattering amplitudes calculated in the $\chi SU(2)$ 
and $\chi SU(3)$ theories, is given by (${\bar M}^2_K= B_0 m_s$):
{\footnotesize \begin{eqnarray}
A_1 \hspace{-2mm}&=&\hspace{-2mm} {1\over 2} + {\bar M}_K^2 \left[ -{{32 L_1^r}\over F_0^2} - {{8 
L_3^r}\over F_0^2} + {{16 L_4^r}\over F_0^2} \right. \nonumber\\
&&\hspace{-2mm}-\left. {5 \over {72 \pi^2 F_0^2}} 
+ {{\sqrt{2} \arctan (\sqrt{2}) }\over 
{27 \pi^2 F_0^2}} + {{7 \ln({4\over 3})}\over{216 \pi^2 F_0^2}} + 
{ 1 \over {32 \pi^2 F_0^2}} \ln {{\bar M}_K^2 \over \mu^2}  \right] + 
O({\bar M}_K^4), \nonumber\\   
\nonumber\\
\sigma_0 \hspace{-2mm}&=&\hspace{-2mm} -{1\over 4} + {\bar M}_K^2 \left[ {{8 L_4^r}\over F_0^2} + 
{{4 L_5^r}\over F_0^2} - {{16 L_6^r}\over F_0^2} - {{8 L_8^r}\over F_0^2} 
\right. \nonumber\\
&&\hspace{-2mm}- \left. {1 \over {288 \pi^2 F_0^2}} - 
{\ln({4\over 3})\over{72 \pi^2 F_0^2}} - {1 \over {72 \pi^2 
F_0^2}} \ln {{\bar M}_K^2 \over \mu^2}  \right] + O({\bar M}_K^4), \nonumber\\
\nonumber\\
A_2 \hspace{-2mm}&=&\hspace{-2mm} -{{32 L_2^r}\over F_0^2} - 
{{8 L_3^r}\over F_0^2} -{2\over {9 \pi^2 F_0^2}} 
+ {{\arctan (\sqrt{2})} \over {27 \sqrt{2} \pi^2 F_0^2}} - {{5 
\ln\left({4\over 3}\right)}\over{108 \pi^2 F_0^2}} + 
{3 \over {16 \pi^2 F_0^2}}\ln {{\bar M}_K^2 \over \mu^2}  + O({\bar M}_K^2), 
\nonumber\\   
\nonumber\\
B_1^r \hspace{-2mm}&=&\hspace{-2mm} -{{4 L_3^r}\over F_0^2} 
-{1\over {576 \pi^2 F_0^2}} - {{5 \arctan (\sqrt{2})}\over {108 \sqrt{2} \pi^2 F_0^2}} - {{31 \ln({4\over 
3})}\over{864 \pi^2 F_0^2}} + {1 \over {96 \pi^2 F_0^2}} 
\ln {{\bar M}_K^2 \over \mu^2}  + O({\bar M}_K^2), \nonumber\\
\nonumber\\
B_2^r \hspace{-2mm}&=&\hspace{-2mm} {13\over {96 \pi^2 F_0^2 {\bar M}_K^2}} + 
{{5 \arctan (\sqrt{2})}\over {72 \sqrt{2} \pi^2 F_0^2 {\bar M}_K^2}} 
+ {{11 \ln({4\over 3})}\over{144 \pi^2 F_0^2 {\bar M}_K^2}} - 
{1 \over {16 \pi^2 F_0^2 {\bar M}_K^0}} 
\ln {{\bar M}_K^2 \over \mu^2}  + 
O({\bar M}_K^2), \nonumber\\
\nonumber\\
B_3^r \hspace{-2mm}&=&\hspace{-2mm} -{L_3^r \over F_0^2} + {L_5^r \over F_0^2} + {5\over {2304 \pi^2 
F_0^2}} + {\arctan (\sqrt{2})\over {108 \sqrt{2} \pi^2 F_0^2}} + {{7 \ln({4\over 
3})}\over{1728 \pi^2 F_0^2}} - {7 \over {768 \pi^2 F_0^2}} 
\ln {{\bar M}_K^2 \over \mu^2} + O({\bar M}_K^2), \nonumber\\
\nonumber \\
C_1^r \hspace{-2mm}&=&\hspace{-2mm} {{32 L_1^r}\over F_0^2} + {{8 L_3^r}\over F_0^2} - {53\over 
{576 \pi^2 F_0^2 }} + {{7 \arctan (\sqrt{2})}\over {108 \sqrt{2} \pi^2 F_0^2 }} + {{11 
\ln({4\over 3})}\over{864 \pi^2 F_0^2 }} - {1 \over {32 \pi^2 F_0^2 }} 
\ln {{\bar M}_K^2 \over \mu^2} + O({\bar M}_K^2) \nonumber\\
\nonumber\\
C_2^r \hspace{-2mm}&-&\hspace{-2mm} C_3^r = {49\over {96 \pi^2 F_0^2 {\bar M}_K^2}} + 
{{29 \arctan (\sqrt{2})}\over {144 \sqrt{2} \pi^2 F_0^2 {\bar M}_K^2}} 
+ {{19 \ln({4\over 3})}\over{72 \pi^2 F_0^2 {\bar M}_K^2}} - 
{3 \over {8 \pi^2 F_0^2 {\bar M}_K^2}} 
\ln {{\bar M}_K^2 \over \mu^2} + 
O({\bar M}_K^0), \nonumber\\
\nonumber \\
C_4^r \hspace{-2mm}&=&\hspace{-2mm} -{59\over {192 \pi^2 F_0^2 {\bar M}_K^4}} - 
{{41 \arctan (\sqrt{2})}\over {576 \sqrt{2} \pi^2 F_0^2 {\bar M}_K^4}} 
- {{13 \ln({4\over 3})}\over{72 \pi^2 F_0^2 {\bar M}_K^4}} + 
{3 \over {8 \pi^2 F_0^2 {\bar M}_K^4}} 
\ln {{\bar M}_K^2 \over \mu^2} + O({\bar M}_K^{-2}), \nonumber \\
\nonumber\\
(C_5^r \hspace{-2mm}&+&\hspace{-2mm} 2C_6^r) + 2(C_8^r +2 C_9^r)+16(C_{13}^r + 
C_{14}^r+2C_{15}^r+C_{16}^r) \nonumber \\
&&\hspace{-2mm}=-{{16 L_1^r}\over F_0^2} - {{4 L_3^r}\over F_0^2} + {{24 L_4^r}\over 
F_0^2}+{{8 L_5^r}\over F_0^2}-{{32 L_6^r}\over F_0^2} - {{16 L_8^r}\over 
F_0^2} \nonumber \\
&&\hspace{3mm}- {1\over {16 \pi^2 F_0^2}} - {{19 \arctan (\sqrt{2})}\over {216 \sqrt{2} 
\pi^2 F_0^2 }} - {{37 \ln({4\over 3})}\over{864 \pi^2 F_0^2 }} + 
{29 \over {576 \pi^2 F_0^2 }} \ln {{\bar M}_K^2 \over \mu^2}  
+ O({\bar M}_K^2), \nonumber \\
\nonumber\\
C_5^r \hspace{-2mm}&+&\hspace{-2mm} 2C_6^r + 4C_7^r +2 C_8^r + 4C_9^r 
=-{{16 L_1^r}\over F_0^2} - {{4 L_3^r}\over F_0^2} + {{24 L_4^r}\over 
F_0^2}+{{4 L_5^r}\over F_0^2} \nonumber \\
&&\hspace{-2mm}-{5\over {32 \pi^2 F_0^2}}
-{{13 \arctan (\sqrt{2})}\over {216 \sqrt{2} \pi^2 F_0^2 }} - {{113 
\ln({4\over 3})}\over{1728 \pi^2 F_0^2 }} + {1 \over {64 \pi^2 F_0^2 }} 
\ln {{\bar M}_K^2 \over \mu^2}  + O({\bar M}_K^2), \nonumber \\
\nonumber\\
-3 \hspace{-3mm}&&\hspace{-6mm} C_3^r - 4( C_{10}^r+2(C_{11}^r + 2C_{12}^r)) + 4 A_2 ( C_5^r +2C_6^r) 
\nonumber \\
&&\hspace{-2mm}=-{13 \over {32 \pi^2 F_0^2 {\bar M}_K^2}} - {{17 \arctan (\sqrt{2})}\over 
{24 \sqrt{2} \pi^2 F_0^2 {\bar M}_K^2}} - {25 \ln({4\over 3})\over 48 \pi^2 
F_0^2 {\bar M}_K^2} + {17 \over 16 \pi^2 F_0^2 {\bar M}_K^2} 
\ln {{\bar M}_K^2 \over \mu^2}  + O({\bar M}_K^0). \nonumber 
\end{eqnarray}}
\unskip The present values for the relevant $SU(3)$ 
constants are given in \cite{Bijnens} at a scale $\mu=M_{\eta}$
\begin{eqnarray}
L_1^r &=& (0.4 \pm 0.3) 10^{-3}, \nonumber\\
L_2^r &=& (1.35 \pm 0.3) 10^{-3},\nonumber\\
L_3^r &=& (-3.5 \pm 1.1) 10^{-3}, \nonumber\\
L_4^r &=& (-0.3 \pm 0.5) 10^{-3}, \nonumber\\
L_5^r &=& (1.4 \pm 0.5) 10^{-3}, \nonumber\\
L_6^r &=& (-0.2 \pm 0.3) 10^{-3}, \nonumber\\
L_8^r &=& (0.9 \pm 0.3) 10^{-3},\nonumber
\end{eqnarray}
which leads to the following numerical estimates
{\small \begin{eqnarray}
A_1 \hspace{-1mm}&=&\hspace{-1mm} (0.7 \pm 0.4) + O({\bar M}_K^4), \nonumber \\
\sigma_0 \hspace{-1mm}&=&\hspace{-1mm} (-0.3 \pm 0.2) + O({\bar M}_K^4), \nonumber \\
A_2 \hspace{-1mm}&=&\hspace{-1mm} (-6.3 \pm 1.5) 10^{-6} \hspace{2mm} {\rm MeV^{-2}} + 
O({\bar M}_K^2), \nonumber \\
B_1^r \hspace{-1mm}&=&\hspace{-1mm} (1.0 \pm 0.5)10^{-6} \hspace{2mm} {\rm MeV^{-2}} + 
O({\bar M}_K^2), \nonumber \\
B_2^r \hspace{-1mm}&=&\hspace{-1mm} (1.1 
\pm 0.0)10^{-11}\hspace{2mm} {\rm MeV}^{-4} + O({\bar M}_K^2), \nonumber \\
B_3^r \hspace{-1mm}&=&\hspace{-1mm} (7.9 
\pm 1.4)10^{-7}\hspace{2mm}{\rm MeV}^{-2} + O({\bar M}_K^2), \nonumber \\
C_1^r \hspace{-1mm}&=&\hspace{-1mm} (-2.0 \pm 
1.5)10^{-6}\hspace{2mm}{\rm MeV}^{-2} + O({\bar M}_K^2), \nonumber \\
C_2^r - C_3^r \hspace{-1mm}&=&\hspace{-1mm} (5.3 \pm 
0.0)10^{-11}\hspace{2mm}{\rm MeV}^{-4} + O({\bar M}_K^0), \nonumber \\
C_4^r \hspace{-1mm}&=&\hspace{-1mm} (-1.6 \pm 
0.0)10^{-16}\hspace{2mm}{\rm MeV}^{-6} + O({\bar M}_K^{-2}) \nonumber 
\end{eqnarray}}
and
{\small \begin{eqnarray}
(C_5^r+2C_6^r)+ 2(C_8^r +2 C_9^r) \nopagebreak &&\nonumber \nopagebreak \\ \nopagebreak
\nopagebreak+16(C_{13}^r + C_{14}^r+2C_{15}^r+C_{16}^r) \hspace{-1mm}&=&\hspace{-1mm} 
(-1.7 \pm 2.1)10^{-6}\hspace{2mm}{\rm MeV}^{-2} + O({\bar M}_K^2), \nonumber \\
C_5^r+2C_6^r + 4C_7^r +2 C_8^r + 4C_9^r 
\hspace{-1mm}&=&\hspace{-1mm} (-2.0 \pm 1.6)10^{-6}\hspace{2mm}{\rm MeV}^{-2} + 
O({\bar M}_K^2), \nonumber \\
- 3 C_3^r - 4 \left[ C_{10}^r+2(C_{11}^r + 2C_{12}^r)\right] && \nonumber\\
+ 4 A_2 ( C_5^r +2C_6^r) \hspace{-1mm}&=&\hspace{-1mm} 
(-1.0 \pm 0.0)10^{-10}\hspace{2mm}{\rm MeV}^{-2} + O({\bar M}_K^0).\nonumber
\end{eqnarray}}
\unskip The quantitative uncertainties are mean square errors of the $L_i$s. These 
could be further reduced by expressing some of these low-energy constants by the 
observables from which they are determined. This was not done. For 
the numerical evaluation, the constants ${\bar M}_K$ and $F_0$ have been replaced 
by their leading-order $\chi SU(3)$ values $M_{\Ka} -1/2 M_{\pi}$ and $F_{\pi}$ 
respectively. This is allowed because these constants only arise in the highest 
significant order. For the masses of the two particles and the pion decay 
constant we have used $M_{\pi}=137.5 \hspace{1mm}{\rm MeV}$, $M_{K}=495.5 \hspace{1mm}{\rm MeV}$ and 
$F_{\pi} = 92.4 \hspace{1mm}{\rm MeV}$.

\section{Relativistic calculation of the scattering amplitude}\label{rel.ap}
In section \ref{scatter} the scattering amplitude is calculated by evaluating 
the fish diagrams in a way that provides a straightforward power-counting 
scheme i.e. the kaon propagator is developed under the integral sign. 
To show that this step is permissible, the scattering 
amplitude is also evaluated in the relativistic framework. In 
the chiral $SU(2)$ theory defined by the general Lagrangian (\ref{Lagrangian}) 
with coupling constants denoted by ${\bar A}_i$, ${\bar B}_i$ and ${\bar C}_i$ 
instead of $A_i$, $B_i$ and $C_i$, fish 
diagrams -- the only problematic diagrams in this context -- are to be 
evaluated in the standard relativistic way \cite{GaLe85}. The result has the 
form of equation (\ref{t32.eqn}) with the low-energy constants replaced by
{\small \begin{eqnarray}
A_1 \hspace{-2mm}&\rightarrow&\hspace{-2mm} {\bar A}^r_1 + \hbar {M^2  \mu^0_{\Ka} \over 3 F^2}, \nonumber \\
A_2 \hspace{-2mm}&\rightarrow&\hspace{-2mm} {\bar A}^r_2 + \hbar \left( -{1 \over 4 \pi^2 F^2} + 
{14  \mu^0_{\Ka} \over 3 F^2} \right), \nonumber \\
\sigma_0 \hspace{-2mm}&\rightarrow&\hspace{-2mm} {\bar \sigma}_0 = {\bar A}_3 + 2{\bar A}_4, \nonumber \\
B^r_1 \hspace{-2mm}&\rightarrow&\hspace{-2mm} {\bar B}^r_1 -\hspace{-0.5mm} \hbar 
\hspace{-1mm} \left[ {1 \over 16 \pi^2 F^2} +  \mu^0_{\Ka} 
\hspace{-0.5mm}\left(-{13 \over 12 F^2} + {5 {\bar A}^r_1 \over 3F^2} - {5 {\bar A}^r_2 M^2 \over 
12 F^2} \right) \hspace{-1mm} \right]\hspace{-0.5mm}, \nonumber \\
B^r_2 \hspace{-2mm}&\rightarrow&\hspace{-2mm} {\bar B}^r_2 +\hspace{-0.5mm} \hbar 
\hspace{-1mm} \left[ {1 \over 8 \pi^2 F^2 M^2} + 
{{\bar A}^r_1 \over 8 \pi^2 F^2 M^2} + { {\bar A}^r_2 \over 8 \pi^2 F^2} 
- \mu^0_{\Ka} \hspace{-0.5mm}\left({4 \over F^2 M^2} + {25 {\bar A}^r_2 \over 12 F^2}  \right) 
\hspace{-1mm} \right]\hspace{-0.5mm}, \nonumber \\
B^r_3 \hspace{-2mm}&\rightarrow&\hspace{-2mm} {\bar B}^r_3 -\hspace{-0.5mm} \hbar 
\hspace{-1mm} \left[ {\sigma_0 \over 16 \pi^2 F^2}-  \mu^0_{\Ka} \hspace{-1mm}\left({1 \over 12 F^2} + {3 \sigma_0 \over 2 F^2} + 
{{\bar A}^r_1 \over 12 F^2} + 
{17 {\bar A}^r_2 M^2 \over 96 F^2}  - {M^2 ({\bar C}^r_5 +2 {\bar C}^r_6) 
\over 2 F^2} \right)\hspace{-1mm} \right] \hspace{-0.5mm}, \nonumber 
\end{eqnarray}}
where
{\small
\[
 \mu^0_{\Ka} = {1\over 32 \pi^2} \ln {B_0 (m_s +{\hat m}) \over \mu^2}.
\] }

The constant $\hbar$, which counts the number of loops, is introduced for 
technical reasons. It helps to ignore two-loop effects which stem from the 
multiplication of the four-point function with the wave-function 
renormalisation constant of the pion.

The relations between $C_i$ and ${\bar C}_i$ have a similar structure but their 
explicit form is of no relevance here. What counts is that the power series 
generated by the development of the kaon propagators converges to the corresponding 
relativistic expressions and that the terms of lower chiral order than indicated 
by the power-counting formula (\ref{chidim}) can be absorbed into the 
low-energy constants without breaking chiral invariance. This would be broken 
if the parameter-free polynomial $\nu$ received a 
contribution from the relativistic calculation, but this does not happen.
\section{Threshold parameters}\label{lepar}

The threshold parameters $b_0^-$, $c_0^-$ and $a_1^-$ can be extracted from the 
$\nu$ odd amplitude (equation \ref{nuodd}) using the definition (\ref{scatlen.def}). The 
parameters $a_0^-$, $b_1^-$ and $a_2^-$ have already been listed in section 
\ref{lep.subsec}.
{\footnotesize \begin{eqnarray}\label{xmin.lec2}
b^-_0 \hspace{-2mm}&=&\hspace{-2mm} {1 \over M_{\pi}(M_{\Ka} + M_{\pi})} {3 M_{\Ka} 
\over 16 \pi F_{\pi}^2} \nonumber \\
&& \hspace{-2mm} + {M_{\pi} \over M_{\Ka} + M_{\pi}} \left( \hspace{-1mm} -{M_{\Ka} \over 384 \pi^3 F_{\pi}^4} + 
{3 \over 16 \pi F_{\pi}^2 M_{\Ka}}- {9 B_1^r M_{\Ka} \over 8 \pi F_{\pi}^2} - {9 B_2^r M_{\Ka}^3 \over 8 \pi 
F_{\pi}^2} + {3 B_3^r M_{\Ka} \over 2 \pi F_{\pi}^2} - {2 M_{\Ka} {\bar \mu}_{\pi} 
\over \pi F_{\pi}^4} \hspace{-1mm} \right) \nonumber \\
&& \hspace{-2mm} + {M_{\pi}^2 \over M_{\Ka} + M_{\pi}} \left( \hspace{-1mm} -{3 \over 64 \pi^3 F_{\pi}^4} - 
{3 B_2^r M_{\Ka}^2 \over 4 \pi F_{\pi}^2} - {3 {\bar \mu}_{\pi} 
\over 2 \pi F_{\pi}^4} \hspace{-1mm} \right), \nonumber \\
&& \hspace{-2mm}  \nonumber \\
c^-_0 \hspace{-2mm}&=&\hspace{-2mm} -{1 \over M_{\pi}^3 (M_{\Ka} + M_{\pi})} {3 M_{\Ka} \over 64 \pi F_{\pi}^2} - 
{1 \over M_{\pi}^2 (M_{\Ka} + M_{\pi})} {3 \over 64 \pi F_{\pi}^2} \nonumber \\
&& \hspace{-2mm} + {1 \over M_{\pi}( M_{\Ka} + M_{\pi})} \left( \hspace{-1mm}{3 
\over 64 \pi F_{\pi}^2 M_{\Ka}} \hspace{-0.5mm}-\hspace{-0.5mm} 
{89 M_{\Ka} \over 1280 \pi^3 F_{\pi}^4} \hspace{-0.5mm}-\hspace{-0.5mm} 
{9 B_1^r M_{\Ka} \over 32 \pi F_{\pi}^2} \hspace{-0.5mm}-\hspace{-0.5mm} 
{9 B_2^r M_{\Ka}^3 \over 32 \pi F_{\pi}^2} \hspace{-0.5mm}-\hspace{-0.5mm} 
{3 B_3^r M_{\Ka} \over 8 \pi F_{\pi}^2} \hspace{-0.5mm}-\hspace{-0.5mm} 
{7 M_{\Ka} {\bar \mu}_{\pi} \over 16 \pi F_{\pi}^4} \hspace{-1mm} \right) \nonumber \\
&& \hspace{-2mm} + {1 \over M_{\Ka} + M_{\pi}} \left( \hspace{-1mm}-{3 \over 64 \pi F_{\pi}^2 M_{\Ka}^2} - 
{239 \over 2304 \pi^3 F_{\pi}^4} + {7 B_1^r \over 32 \pi F_{\pi}^2} - {21 B_2^r M_{\Ka}^2 \over 32 \pi 
F_{\pi}^2} - {3 B_3^r \over 8 \pi F_{\pi}^2} - {65 {\bar \mu}_{\pi} 
\over 48 \pi F_{\pi}^4} \hspace{-1mm} \right), \nonumber \\
&& \hspace{-2mm}  \nonumber \\
a^-_1 \hspace{-2mm}&=&\hspace{-2mm} {1 \over M_{\Ka} + M_{\pi}} {1 \over 16 \pi F_{\pi}^2} + 
{M_{\pi} \over M_{\Ka} + M_{\pi}} \left( \hspace{-1mm} -{5 M_{\Ka} \over 1152 \pi^3 F_{\pi}^4} + 
{B_1^r M_{\Ka} \over 4 \pi F_{\pi}^2} - {M_{\Ka} {\bar \mu}_{\pi} \over 12 \pi 
F_{\pi}^4} \hspace{-1mm} \right) \nonumber \\
&& \hspace{-2mm} + {M_{\pi}^2 \over M_{\Ka} + M_{\pi}} \left( \hspace{-1mm} -{1 \over 192 \pi^3 F_{\pi}^4} 
- {B_1^r \over 8 \pi F_{\pi}^2} - {3 B_2^r M_{\Ka}^2 \over 8 \pi 
F_{\pi}^2} + {B_3^r \over 2 \pi F_{\pi}^2} - {3 {\bar \mu}_{\pi} 
\over 4 \pi F_{\pi}^4} \hspace{-1mm} \right).
\end{eqnarray}}

As explained in section \ref{lep.subsec}, the constants $B_1^r$ and $B_2^r$ 
can be eliminated in favour of the observables $a_2^-$ and $b_1^-$. The result 
is
{\footnotesize \begin{eqnarray}\label{aba2}
b^-_0 \hspace{-2mm}&=&\hspace{-2mm}  {1 \over M_{\pi}(M_{\Ka} + M_{\pi})} 
\left( {3 M_{\Ka} \over 16 \pi F_{\pi}^2} \hspace{-0.5mm}+\hspace{-0.5mm} 
{M_{\Ka}^3 \over 640 \pi^3 F_{\pi}^4} \right) \hspace{-0.5mm}+\hspace{-0.5mm} 
{1 \over M_{\Ka} + M_{\pi}} \left( {3 \over 32 \pi F_{\pi}^2} \hspace{-0.5mm}+\hspace{-0.5mm} 
{7 M_{\Ka}^2 \over 960 \pi^3 F_{\pi}^4} \hspace{-0.5mm}-\hspace{-0.5mm} 
{ 15 M_{\Ka}^3 a_2^- \over 2} \right) \nonumber \\
&&\hspace{-2mm}+ {M_{\pi} \over M_{\Ka} + M_{\pi}} \left( {1 \over 4 \pi M_{\Ka} F_{\pi}^2} + {3 
B^r_3 M_{\Ka} \over 2 \pi F_{\pi}^2} + {123 M_{\Ka} \over 1280 \pi^3 
F_{\pi}^4} - {M_{\Ka} {\bar \mu}_{\pi} \over 8 \pi F_{\pi}^4} - {55 M_{\Ka}^2 a_2^- \over 
2} +3 M_{\Ka}^2 b_1^- \right) \nonumber \\
&&\hspace{-2mm}+ {M_{\pi}^2 \over M_{\Ka} + M_{\pi}} \left( {1 \over 32 \pi^3 F_{\pi}^4} - 15 
M_{\Ka} a_2^- +5 M_{\Ka} b_1^- \right), \nonumber \\
&& \nonumber \\
a^-_1 \hspace{-2mm}&=&\hspace{-2mm}  {1 \over M_{\Ka} + M_{\pi}} 
\left( {1 \over 16 \pi F_{\pi}^2} - {M_{\Ka}^2 \over 1920 \pi^3 F_{\pi}^4} 
\right) \hspace{-0.5mm}+\hspace{-0.5mm} {M_{\pi} \over M_{\Ka} + M_{\pi}} \left( 
{1 \over 32 \pi M_{\Ka} F_{\pi}^2} \hspace{-0.5mm}+\hspace{-0.5mm} 
{M_{\Ka} \over 960 \pi^3 F_{\pi}^4} \hspace{-0.5mm}+\hspace{-0.5mm} 
{5 M_{\Ka}^2 a_2^- \over 2 } \right) \nonumber \\
&&\hspace{-2mm}+ {M_{\pi}^2 \over M_{\Ka} + M_{\pi}} \left( {B^r_3 \over 2 \pi F_{\pi}^2 } + 
{73 \over 2304 \pi^3 F_{\pi}^4} - {{\bar \mu}_{\pi} \over 24 \pi F_{\pi}^4} + {5 M_{\Ka} a_2^- \over 2} + 
M_{\Ka} b_1^- \right). 
\end{eqnarray}}

The only input from $\chi SU(3)$ is the constant $B^r_3=7.9 \cdot 10^{-7} 
\hspace{1mm}{\rm MeV}^{-2}$ which appears only in $\chi SU(2)$-suppressed terms. 

Similarly, the threshold parameters $a_0^+$, $b_0^+$, $c_0^+$, $a_1^+$, 
$b_1^+$ and $a_2^+$ can be extracted from the $\nu$-even amplitude defined by 
equation (\ref{nuodd}):
{\footnotesize \begin{eqnarray}\label{ahoch+}
a^+_0 \hspace{-2mm}&=&\hspace{-2mm}  {M_{\pi}^2 \over M_{\Ka} + M_{\pi}} \left( -{3 A_1 \over 8 \pi F_{\pi}^2} 
- {3 A_2 M_{\Ka}^2 \over 8 \pi F_{\pi}^2} - {3 \rho \over 4 \pi F_{\pi}^2} 
\right) \nonumber \\
&& \hspace{-2mm} + {M_{\pi}^4 \over M_{\Ka} + M_{\pi}} \left( {96 \lambda_{(\nu^4)} M_{\Ka}^4 \over \pi} 
+ {6 \lambda_{(\nu^2 m^2)} M_{\Ka}^2 \over \pi} + {3 \lambda_{(m^4)} \over 8 \pi} 
\right), \nonumber \\
&& \nonumber \\
b^+_0 \hspace{-2mm}&=&\hspace{-2mm}  {1 \over M_{\Ka} + M_{\pi}} \left( -{3 A_1 \over 8 \pi F_{\pi}^2} 
- {3 A_2 M_{\Ka}^2 \over 8 \pi F_{\pi}^2} \right) 
+ {M_{\pi} \over M_{\Ka} + M_{\pi}} \left( {3 A_1 \over 16 \pi F_{\pi}^2 M_{\Ka}} 
- {3 A_2 M_{\Ka} \over 16 \pi F_{\pi}^2} + {3 \rho \over 8 \pi F_{\pi}^2 M_{\Ka}} 
\right) \nonumber \\
&& \hspace{-2mm} + {M_{\pi}^2 \over M_{\Ka} + M_{\pi}} \left( {3 \over 16 \pi^3 F_{\pi}^4} - 
{\rho \over 128 \pi^3 F_{\pi}^4} - 
{A_1 \over 256 \pi^3 F_{\pi}^4} -{A_2 M_{\Ka}^2 \over 768 \pi^3 F_{\pi}^4} - 
{3 A_2 \over 8 \pi F_{\pi}^2} \right. \nonumber \\
&& \hspace{+2cm} + \left. {192 \lambda_{(\nu^4)} M_{\Ka}^4 \over \pi} - {12 \lambda_{(\nu^2 t)} M_{\Ka}^2 \over 
\pi} + {6 \lambda_{(\nu^2 m^2)} M_{\Ka}^2 \over \pi} - {3 \lambda_{(t m^2)} \over 4 
\pi} \right), \nonumber \\
&& \nonumber \\
c^+_0 \hspace{-2mm}&=&\hspace{-2mm}  {1 \over M_{\pi}(M_{\Ka} + M_{\pi})} \left( {9 A_1 \over 64 \pi F_{\pi}^2 
M_{\Ka}} - {3 A_2 M_{\Ka} \over 64 \pi F_{\pi}^2} - {3 \rho \over 32 \pi F_{\pi}^2 M_{\Ka}} 
\right) \nonumber \\
&& \hspace{-2mm} + {1 \over M_{\Ka} + M_{\pi}} \left( -{3 A_1 \over 64 \pi F_{\pi}^2 M_{\Ka}^2} -{23 
A_2 \over 64 \pi F_{\pi}^2} - {3 \rho \over 32 \pi F_{\pi}^2 M_{\Ka}^2} + {96 \lambda_{(\nu^4)} M_{\Ka}^4 \over \pi} - 
{12 \lambda_{(\nu^2 t)} M_{\Ka}^2 \over 
\pi} + {2 \lambda_{(t^2)} \over \pi} \right. \nonumber \\
&& \hspace{+2cm} + \left. {3 \over 16 \pi^3 F_{\pi}^4} - {A_1 \over 40 \pi^3 F_{\pi}^4} - 
{43 A_2 M_{\Ka}^2 \over 5760 \pi^3 F_{\pi}^4} - {19 \rho \over 480 \pi^3 F_{\pi}^4} 
\right), \nonumber \\
&& \nonumber \\
a^+_1 \hspace{-2mm}&=&\hspace{-2mm}  {1 \over M_{\Ka} + M_{\pi}} {A_1 \over 8 \pi F_{\pi}^2} - 
{M_{\pi} \over M_{\Ka} + M_{\pi}} {A_2 M_{\Ka} \over 8 \pi F_{\pi}^2} \nonumber \\
&& \hspace{-2mm} + {M_{\pi}^2 \over M_{\Ka} + M_{\pi}} \left( {4 \lambda_{(\nu^2 t)} M_{\Ka}^2 \over \pi} 
+ {\lambda_{(t m^2)} \over 4 \pi} + {1 \over 16 \pi^3 F_{\pi}^4} + {A_1 \over 768 \pi^3 F_{\pi}^4} + 
{A_2 M_{\Ka}^2 \over 2304 \pi^3 F_{\pi}^4} + {\rho \over 384 \pi^3 F_{\pi}^4} \right), 
\nonumber \\
b^+_1 \hspace{-2mm}&=&\hspace{-2mm}  {1 \over M_{\pi}(M_{\Ka} + M_{\pi})} \left( -{A_1 \over 16 \pi F_{\pi}^2 
M_{\Ka}} - {A_2 M_{\Ka} \over 16 \pi F_{\pi}^2} \right) \nonumber \\
&& \hspace{-2mm} + {1 \over M_{\Ka} + M_{\pi}} \left( {4 \lambda_{(\nu^2 t)} M_{\Ka}^2 \over \pi} 
- {\lambda_{(t^2)} \over \pi} + {1 \over 8 \pi^3 F_{\pi}^4} + {A_1 \over 80 \pi^3 F_{\pi}^4} + 
{43 A_2 M_{\Ka}^2 \over 11520 \pi^3 F_{\pi}^4} + {19 \rho \over 960 \pi^3 F_{\pi}^4} 
 \right), \nonumber \\
&& \nonumber \\
a^+_2 \hspace{-2mm}&=&\hspace{-2mm}   {1 \over M_{\Ka} + M_{\pi}} \left( -{A_2 \over 80 \pi F_{\pi}^2} + 
{\lambda_{(t^2)} \over 5 \pi} - {A_1 \over 400 \pi^3 F_{\pi}^4} - {43 A_2 M_{\Ka}^2 \over 57600 \pi^3 
F_{\pi}^4} - {19 \rho \over 4800 \pi^3 F_{\pi}^4} \right). 
\end{eqnarray}}

\pagebreak
\end{document}